\documentclass[preprint,preprintnumbers,amsmath,amssymb]{revtex4}
\usepackage{amssymb,latexsym,amsmath}

\usepackage[activeacute,english]{babel}
\usepackage{fancyhdr}
\usepackage{textcomp}
\usepackage{amsfonts}
\usepackage{graphicx}
\usepackage{graphics}
\usepackage{tocbibind}
\usepackage[toc,page]{appendix}

\usepackage[active]{srcltx}

\begin{document}

\title{Helium-like ions in $d$-dimensions: analyticity and generalized ground state Majorana solutions}

\author{Adrian M. Escobar-Ruiz}
\email{admau@xanum.uam.mx}

\author{Horacio Olivares-Pilón}
\email{horop@xanum.unam.mx}

\author{Norberto Aquino}
\email{naa@xanum.uam.mx}

\author{Salvador A. Cruz}
\email{cruz@xanum.uam.mx}

\affiliation{Departamento de Fisica, Universidad Aut\'onoma Metropolitana-Iztapalapa,
San Rafael Atlixco 186, C.P. 09340 Ciudad de M\'exico, M\'exico}

\vspace{2.0cm}

\begin{abstract}
Non-relativistic Helium-like ions $(-e,-e,Ze)$ with static nucleus in a $d-$dimensional space $\mathbb{R}^d$ ($d>1$) are considered. Assuming $r^{-1}$ Coulomb interactions, a 2-parametric correlated Hylleraas-type trial function is used to calculate the ground state energy of the system in the domain $Z \leq 10$. For odd $d=3,5$, the variational energy is given by a rational algebraic function of the variational parameters whilst for even $d=2,4$ it is shown for the first time that it corresponds to a more complicated non-algebraic expression. This twofold analyticity will hold for any $d$. It allows us to construct reasonably accurate approximate solutions for the ground state energy $E_0(Z,d)$ in the form of compact analytical expressions. We call them generalized Majorana solutions. They reproduce the first leading terms in the celebrated $\frac{1}{Z}$ expansion, and serve as generating functions for certain correlation-dependent properties. The (first) critical charge $Z_{\rm c}$ \textit{vs} $d$ and the Shannon entropy $S_{r}^{(d)}$ \textit{vs} $Z$ are also calculated within the present variational approach. In the light of these results, for the physically important case $d=3$ a more general 3-parametric correlated Hylleraas-type trial is used to compute the finite mass effects in the Majorana solution for a three-body Coulomb system with arbitrary charges and masses. It admits a straightforward generalization to any $d$ as well. Concrete results for the systems $e^-\,e^-\,e^+$, $H_2^+$ and $H^-$ are indicated explicitly. Our variational analytical results are in excellent agreement with the exact numerical values reported in the literature.
\end{abstract}

\maketitle

\hskip 1cm

\section{Introduction}

Ever since the advent of quantum mechanics the description of the helium atom and isoelectronic ions, which illustrate the properties of few-electron atomic ions in non-relativistic quantum electrodynamics (QED), has attracted the interest of physicists and chemists. Solving approximately this well known system is one of the basic problems in nonrelativistic quantum mechanics: it appears in many textbooks. Moreover, the helium atom continues to play a key role as a testing ground for approximate methods, being the electron-electron repulsion
term the cause of non-separability of the Schrödinger equation, and especially for techniques to assess the uncertainty.

\vspace{0.2cm}

In fact, as early as 1929, Hylleraas \cite{Hylleraas} studied the ground state of the He atom using a very simple correlated trial wavefunction $\psi=\psi(r_1, r_2, r_{12})$ in terms of all three relative mutual distances among particles. Such an early attempt lead to a more general approach often referred nowadays under the name of Hylleraas method (see the review article \cite{DrakeG}).

Also, as it is now recognized, during the same time period Ettore Majorana had advanced
work in this direction appearing in his unpublished notes ( see \cite{Esposito} for historical account and
further references). In that work Majorana proposed, for three-dimensional He-like ions, an
empirical one-parameter formula for the ground-state energy versus nuclear charge $Z$ (in
a.u.):
\begin{equation}\label{Majorana}
  E  \ = \ -Z^2 \ + \ \frac{5}{8}\,Z \ + \ b \ ,
\end{equation}

where $b$ is a parameter, which can be found variationally. In fact, the analytical expression (\ref{Majorana}) corresponds to the variational ground state energy with non-correlated Hylleraas function $e^{-Z\,(r_1+r_2)}$. We call this the 3D Majorana solution.

\vspace{0.2cm}

Since then more accurate treatments of He-like systems in three dimensions have been
reported, such as the use of a modified Hylleraas-like function in the form of
exponentials in all three interparticle distances, employed originally in variational
calculations by Delves and Kalotas \cite{Delves} and then by Thakkar and Smith \cite{Thakkar} to compute
the ground state energy of He-like ions. Further efforts to obtain higher accurate
calculations have been developed. Notably, one of the most accurate calculations reported so
far ($\sim35$ significant digits) for the ground-state energy of He-like systems with static nuclear
charges $Z = 1–10$ have been reported \cite{Nakashima} and efforts to improve this accuracy still
continue \cite{Drake-Zong-Chao Yan,Korobov2}. Also, a variational strategy using a sequence of Hylleraas-type basis
functions has been put forward in the study of the general three-body Coulomb
problem for arbitrary particle masses and charges \cite{Frolov, Frolov2}. All of these studies
quantitatively reproduced the energy and other physical properties, particularly of He-like
atomic ions. However, no explicit approximate expression for the ground state
energy $E(Z)$ is given. Moreover, the wavefunctions used in such calculations require
of a large number of terms (some of them up to thousands) hence hindering a clear
understanding of their physical features.

\vspace{0.2cm}

So far, the pioneering work by Hylleraas and Majorana for the variational
treatment of the three-dimensional He-like system has not only been of hallmark importance for
further refinements in the accurate treatment of three-body Coulomb systems, but as demonstrated in \cite{TVH} at $d=3$ an Hylleraas-type function for the ground state is one of the simplest exact solutions of the effective theory which reproduces non-relativistic quantum electrodynamics effects. Also, accurate knowledge of the ground-state energy allows to find non-relativistic QED effects and, perturbatively,
the relativistic QED correction \cite{TVH}. In this connection, the interested reader is kindly
addressed to the review paper \cite{Drake 2020} for a recent account of progress in assessing the
accuracy of atomic and molecular data regarding theoretical or computational work.

\vspace{0.2cm}

We now ask how appropriate is the Hylleraas-Majorana solution in $d-$dimensions for
the study of He-like ions and, in general, of three-body Coulomb systems. Recently \cite{TME3-d,Willard:2018} several aspects of a general quantum three-body system in $d$-dimensions $(d>1)$ with arbitrary masses have been reported (see also \cite{Gu}). Needless to
say, the $d-$dimensional He atom has been analyzed in some detail by several authors \cite{Herrick}-\cite{Herschbach} mainly within the infinite nuclear mass approximation. Again, all these treatments require of elaborate theoretical and numerical procedures which make it difficult to follow a clear physical picture of the role of the relevant parameters involved. Hence, it is desirable to search for compact –yet reasonably accurate– physical ground state wavefunctions in $d$-dimensions for these systems. This is the main focus of this work as explained below.

\vspace{0.2cm}

In the present work we aim to improve the 3D - Majorana solution (\ref{Majorana}) using a single
fully correlated function $e^{-\alpha \,Z\,(r_1+r_2)-\beta\,r_{12}}$ and, mainly, to extend it to de $d$-dimensional case.
Additionally, we seek for analyticity, namely a compact analytical formula (a
generalized Majorana solution) for the ground state for any $d >1$. Our goal is three-fold:
firstly, to describe in a unified way a simple analytical method for calculating - with a
reasonable accuracy - the ground state energy of a d-dimensional non-relativistic He-like
ion with fixed nucleus. This may be considered as a natural generalization of the
3D-Majorana solution to any dimension. Secondly, to show that this method can be
easily improved and generalized to the $d$-dimensional case of a three-body Coulomb
system with arbitrary masses and charges. This is explicitly shown for three dimensions. Thirdly, with this compact representation, two characteristics for $d=2,3,4,5$ and $Z \in [2,10]$ are calculated: the critical charge $Z_c$ and the Shannon entropy $S_{r}^{(d)}$ (see below).

\vspace{0.2cm}

Again, it is worth pointing out here that the above mentioned generalized Majorana formula
obtained in this work has an analytical expression and provides a very good accuracy. In particular, a novel non-liner change of variables in $r$-space $(r_1,r_2, r_{12})$ is introduced that allows to evaluate explicitly the variational energy for $d=2,4,...,2N$. To the best of our knowledge, the corresponding 2D Majorana solution is presented for the first time.

\vspace{0.2cm}

The structure of the paper is as follows. In Section II the general setting of the problem is explained. For the lowest values $d=2,3,4,5$ the corresponding concrete solutions, the generalized Majorana solutions, are derived in the Sec. III. Agreement of those results with highly accurate numerical values reported in the literature is presented. Somewhat unexpectedly, we also found that the case of $d$ odd allows an exact analytic solution in terms of elementary functions. Would this property be true in four- and five-body atomic and molecular systems: it is interesting open question. The expansion of the variational results obtained for large $Z$ is compared with the celebrated perturbative expansion $1/Z$ for the ground state energy. The next Sec. IV treats two additional important properties of the system, the Shannon entropy and the (first) critical charge. Finally, in Sec. V we construct explicitly at $d=3$ an improved Majorana solution by direct evaluation of the expectation value of the energy using a more general correlated trial function. It leads to an analytic answer for arbitrary values of the charges and masses. While the form of the reduced Hamiltonian changes, the problem remains 3-dimensional, leading to a rather simple analytic answer which can be immediately generalized to higher values of $d$. This section contains discussion of possible applications to other exotic three-body systems as well. Atomic units $(e=1,\hbar=1)$ are used throughout unless otherwise specified.

\section{Generalities}
\label{}

Assuming $r^{-1}$ Coulomb interactions, the Hamiltonian of a $d$-dimensional Helium-like ion with static nuclei $Z>0$ takes the form:
\begin{equation}
{\cal H}\ = \    -\frac{1}{2}\,\Delta^{(d)}_1\  - \ \frac{1}{2}\Delta^{(d)}_2 \ - \ \frac{Z}{r_{1}} \ - \ \frac{Z}{r_{2}}\ + \ \frac{1}{r_{12}} \ ,
\label{H}
\end{equation}
where $r_{1(2)}$ is the distance from nuclear charge $Z$ to $i$th electron, $r_{12}$ is the inter-electronic distance, and $\Delta^{(d)}_{1(2)}$ is $d$-dimensional Laplacian associated with $i$th electron. The eigenvalue problem for ${\cal H}$ is defined on the configuration space ${\mathbb R}^{2d}$.

Since the potential depends on the Hylleraas coordinates $r_1, r_2$ and $r_{12}$ only, it seems natural to seek square integrable eigenfunctions of (\ref{H}), ${\cal H}\,\psi=E\,\psi$, in the form:

\begin{equation}\label{psiH3}
  \psi \ = \ \psi(\,r_1,\,r_2,\,r_{12}  \,) \ .
\end{equation}

The ground state is of this type. It can be shown that these solutions (\ref{psiH3}) with zero total angular momentum are described by a reduced Hamiltonian operator ${\cal H}_r$. Explicitly \cite{TME3-d}
\begin{equation}\label{Hr}
  {\cal H}_r \ = \ -\frac{1}{2}\Delta_{r}^{(3)}(r_1,\,r_2,\,r_{12})  \ - \ \frac{Z}{r_{1}} \ - \ \frac{Z}{r_{2}}\ + \ \frac{1}{r_{12}}\ ,
\end{equation}
where
\begin{equation}
\begin{split}
  \Delta_{r}^{(3)}\  =\  & \frac{1}{r_{1}^{d-1}}\frac{\partial}{\partial r_{1}}\left(r_{1}^{d-1}\frac{\partial}{\partial r_{1}}\right)\ + \ \frac{1}{r_{2}^{d-1}}\frac{\partial}{\partial r_{2}}\left(r_{2}^{d-1}\frac{\partial}{\partial r_{2}}\right)\ +\ \frac{2}{r_{12}^{d-1}}\frac{\partial}{\partial r_{12}}\left(r_{12}^{d-1}\frac{\partial}{\partial r_{12}}\right)
  \\ &
  \ + \frac{r_{1}^{2}+r_{12}^{2}-r_{2}^{2}}{r_{1}\,r_{12}}\,\frac{\partial}{\partial r_{12}}\left(\frac{\partial}{\partial r_{1}}\right)\ +\  \frac{r_{1}^{2}+r_{12}^{2}-r_{1}^{2}}{r_{2}\,r_{12}}\,\frac{\partial}{\partial r_{12}}\left(\frac{\partial}{\partial r_{2}}\right) \ .
\end{split}
\end{equation}
The Hamiltonian (\ref{Hr}) describes a three-dimensional particle moving in a curved space. As mentioned above, it describes the angle-independent solutions ($\cal S$-states) of the original Hamiltonian (\ref{H}). The corresponding configuration space is given by $\Omega \,\subset\, {\mathbb R}_+^{3}$. The ranges of the three variables $r_{ij}$ are coupled in that they must satisfy a “triangle condition”: their lengths must be such that they can form a triangle.

The operator (\ref{Hr}) is essentially self-adjoint with respect to the radial measure

\begin{equation}\label{rm}
dV \ = \ \frac{2^d\,\pi^{d-1}}{(d-2)!}\,r_{1}\,r_{2}\,r_{12}\,S^{d-3}\,dr_{1}\,dr_{2}\,dr_{12} \qquad , \qquad \ (d>1) \ ,
\end{equation}

\[
S \ = \ \frac{1}{4}\sqrt{(r_1 \,+\,r_2\,+\,r_{12})(r_1 \,+\,r_2\,-\,r_{12})(r_1 \,-\,r_2\,+\,r_{12})(-r_1 \,+\,r_2\,+\,r_{12})} \ ,
\]

$S$ is the area of the triangle formed by the two electrons and the charge $Z$. The volume element (\ref{rm}) is greatly simplified at $d=3$, $dV=8\,\pi^2\,r_{1}\,r_{2}\,r_{12}\,dr_{1}\,dr_{2}\,dr_{12}$.

In $dV = w(r_{1},\,r_{2},\,r_{12};\,d)\,dr_{1}\,dr_{2}\,dr_{12}$ the weight function $w \propto r_{1}\,r_{2}\,r_{12}\,S^{d-3}$ for odd $d\geq 3$ becomes a polynomial $w \rightarrow P_N(r_{1},\,r_{2},\,r_{12})$ of degree $N=2\,d-3$ whilst at even $d\geq 2$ it is an irrational algebraic expression.

The expectation value of the Hamiltonian (\ref{Hr}), the energy functional, is given by

\begin{equation}\label{Evar0}
  E_{\rm var}[\psi,\,d] \ = \ \frac{\int_{0}^{\infty}\int_{0}^{\infty}\int_{|r_1-r_2|}^{r_1+r_2}\, (\psi^\ast\,{\cal H}_r\,\psi)\,w\,dr_{1}\,dr_{2}\,dr_{12}}{\int_{0}^{\infty}\int_{0}^{\infty}\int_{|r_1-r_2|}^{r_1+r_2}\, {\mid \psi \mid}^2\,w\,dr_{1}\,dr_{2}\,dr_{12}} \ .
\end{equation}

A considerable simplification results from going over to the so-called perimetric coordinates used by Pekeris in his helium calculations \cite{PEKERIS}. These perimetric coordinates are defined by the linear relations
\begin{equation}\label{perimetric}
\begin{aligned}
& \eta \ = \ -r_1 \ + \ r_2 \ + \ r_{12}
\\ &
\sigma \ = \ r_1 \ - \ r_2 \ + \ r_{12}
\\ &
\tau \ = \ 2\,(r_1 \ + \ r_2 \ - \ r_{12})
\\ &
dr_{1}\,dr_{2}\,dr_{12} \ = \ \frac{1}{8} d\eta\,d\sigma\,d\tau \ .
\end{aligned}
\end{equation}

By the above transformation the limits of the three integrals in $E_{\rm var}$ (\ref{Evar0}) become independent of
each other

\begin{equation}\label{Evar}
  E_{\rm var}[\psi,\,d] \ = \ \frac{\int_{0}^{\infty}\int_{0}^{\infty}\int_{0}^{\infty}\, (\psi^\ast\,{\cal H}_r\,\psi)\,w\,d\eta\,d\sigma\,d\tau}{\int_{0}^{\infty}\int_{0}^{\infty}\int_{0}^{\infty}\, {\mid \psi \mid}^2\,w\,d\eta\,d\sigma\,d\tau} \ .
\end{equation}

\bigskip

Now, for the ground state and arbitrary $d$ let us take the celebrated Hylleraas-type trial function
\begin{equation}\label{psi1}
  \psi_{\small \rm Hs} \ = \ e^{-\alpha\,Z\,(r_1\,+\,r_2)\,-\,\beta\,r_{12} }  \ = \ e^{-\frac{\alpha\,Z}{2}  (\eta +\sigma +\tau )\,-\,\frac{\beta}{2} (\eta \,+\,\sigma )} \ ,
\end{equation}

where $\alpha>0$ and $\beta \leq 0$ are non-linear variational parameters. They define the electron-nuclei and the electron-electron cusp parameters, respectively. Hence, they measure the local quality of the approximate wave function near the Coulomb singularities. The one term function (\ref{psi1}) incorporates the electronic correlation explicitly via the exponent $\beta\,r_{12}$.

For actual calculations, we further introduce a new nonlinear change of variables in the $r-$space given by
\begin{equation}\label{uvar}
\begin{aligned}
& u_1 \ = \ \eta \ + \ \sigma \ = \ r_{12} \ - \  3\,(\, r_{2}\,-\,r_1\,)
\\ &
u_2 \ = \ \eta \,\sigma \ = \ 2 \,(\,r_1 \, - \, r_2\,) (\,r_1 \,+\, r_{12} \,-\,r_2\,)
\\ &
u_3 \ = \ \tau \ = \ 2\,(r_1 \ + \ r_2 \ - \ r_{12})
\\ &
dr_{1}\,dr_{2}\,dr_{12} \ = \ \frac{1}{8} d\eta\,d\sigma\,d\tau \ = \  \frac{1}{2}\frac{1}{\sqrt{u_1^2\,-\,4\,u_2}} du_1\,du_2\,du_3     \ .
\end{aligned}
\end{equation}

It turns out that for even $d$, unlike the perimetric coordinates (\ref{perimetric}), these $u-$variables (\ref{uvar}) allow us to evaluate the triple integral $E_{\rm var}$ (\ref{Evar0}) analytically in closed form. To the best of our knowledge this set of $u$-coordinates is presented for the first time.

Note that explicit analytical formulas for the expectation value $\langle P\rangle_{\small  \psi_{\small \rm Hs}}$ of any two-variable polynomial function $P=P(r_1+r_2,r_{12})$ can be derived in a simple manner as well. To this end, let us take the denominator in $E_{\rm var}[\psi_{\small \rm Hs},\,d]$ (\ref{Evar})
\begin{equation}\label{Lambd}
 \Lambda(\alpha,\,\beta) \ = \ {\rm Den} \,E_{\rm var} \ ,
\end{equation}

which can be evaluated exactly. From (\ref{Evar}), we immediately obtain the relation \cite{Calais}

\begin{equation}\label{relat}
  \langle \, P(r_1+r_2,r_{12})\, \rangle \ = \   \frac{1}{\Lambda(\alpha,\,\beta)} P\bigg(-\frac{1}{2\,Z}\partial_{\alpha},\,-\frac{1}{2}\partial_{\beta}\bigg)\,\Lambda(\alpha,\,\beta) \ ,
\end{equation}

where $\partial_{\alpha} \equiv \frac{\partial}{\partial \,\alpha},\,\partial_{\beta} \equiv \frac{\partial}{\partial \,\beta}$. It implies that $E_{\rm var}$ plays the role of a generating function.

\section{Generalized ground state Majorana solutions: concrete results}

In this Section, we present the generalized Majorana solutions (GMS) for the lowest values of $d=2,3,4,5$. They are obtained from the exponential correlated Hylleraas-type trial function $\psi_{\small \rm Hs} =  e^{-\alpha\,Z\,(r_1\,+\,r_2)\,-\,\beta\,r_{12} } $ (\ref{psi1}).

\subsection{Case $d=3$}

Let us start with the most relevant physical situation $d=3$. In this case, the factor $w$ appearing in the functional energy $E_{\rm var}$ (\ref{Evar}) is a cubic polynomial in perimetric coordinates (\ref{perimetric}), $w(d=3) \propto (\eta +\sigma ) (2 \,\eta +\tau ) (2 \,\sigma +\tau )$. We immediately obtain the variational energy with respect to (\ref{psi1})

\begin{equation}\label{Evar2}
E_{\rm var}^{(3D)} \ = \ \frac{(\beta +\alpha  Z) \left(\beta ^3+\beta ^2+8 \alpha ^3 Z^3-16 \alpha ^2 Z^3+7 \alpha ^2 \beta  Z^2+5 \alpha ^2 Z^2-4 \alpha  \beta  Z^2+4 \alpha  \beta ^2 Z+4 \alpha  \beta  Z\right)}{\beta ^2+8 \alpha ^2 Z^2+5 \alpha  \beta  Z}\ ,
\end{equation}

hence, it is a rational function of the non-linear parameters $\alpha$ and $\beta$. At $\beta=0$, the analytical expression (\ref{Evar2}) reduces to
\begin{equation}\label{E3Dal}
E_{\rm var}^{(3D)}(\beta=0) \ = \   (\alpha -2)\, \alpha \, Z^2 \ + \ \frac{5\, \alpha\,  Z}{8}\ .
\end{equation}
At $\alpha=1,\,\beta=0$, it further simplifies to the Majorana expression (\ref{Majorana}) with $b=0$
\begin{equation}\label{}
E_{\rm var}^{(3D)}(\alpha=1,\,\beta=0) \ = \   -Z^2 \ + \ \frac{5}{8}\,Z\ .
\end{equation}
Moreover, as a result of the minimization procedure the optimal parameters $\alpha$ and $\beta$ turn out to be almost constant in the domain $2\leq Z \leq 10$, see Fig. \ref{varp}. Thus, at $\alpha=\alpha_{\rm average}$ and $\beta=\beta_{\rm average}$\footnote{We denote by $\alpha_{\rm average}$ the average value of the optimal parameter $\alpha$ in the region $2\leq Z \leq10$.}, we call (\ref{Evar2}) generalized 3D Majorana solution. In particular, for the ground state of the Helium atom ($Z=2$) the minimization of (\ref{Evar2}) gives the result
\[
E_{0}^{(3D)}(Z=2) \ = \ -2.889618\,a.u.
\]
at $\alpha=0.929044,\,\beta=-0.254746$, in complete agreement with \cite{Aquino}. Hence, a nuclear-electron cusp $\nu_1=-1.858$ and an electron-electron cusp $\nu_2=0.2547$ . They have to be compared with the \emph{exact} results by Schwartz \cite{Schwartz}:
\[
E_{\rm exact}^{(3D)}(Z=2) \ = \ -2.903724\,a.u. \qquad , \qquad \nu_1^{\rm(exact)} \ = \ -2  \qquad , \qquad \nu_2^{\rm (exact)} \ = \ \frac{1}{2} \ .
\]

The difference in the energy of $0.014$ a.u. $\sim 0.380$eV is the actual error. For comparison, $ k_B\,T\sim 0.026$ eV at room temperature. Note that in the Hartree–Fock approximation, $\beta=0$, the error ($\sim 1.14$eV) is three times larger. One might say that all of chemistry is buried in the correlation energy, and that is why it is so important to get it right.

It is evident that the two-parametric trial function (\ref{psi1}), which provides rather accurate results, can be improved even without breaking the aforementioned analyticity of $E_{\rm var}$. As mentioned in the Introduction such a program has been realized at $d=3$ (see also \cite{Harris}). In order to keep the calculations as simple as possible we choose (\ref{psi1}) to focus on the conceptual part of the present consideration.

\subsubsection{$\frac{1}{Z}$ and Puiseux expansions }

In this subsection, in order to estimate the quality of the 3D GMS we will check its compatibility with well-established results found from perturbation theory.

Firstly, it is known that at large $Z$ the ground state energy can be estimated by the $1/Z$ perturbative expansion \cite{Kato:1980},
\begin{equation}
\label{1overZ}
  E(Z)\ =\ -B_0\, Z^2 \ + \  B_1\, Z\  + \ B_2 \ + \  O\bigg(\frac{1}{Z}\bigg)\ ,
\end{equation}
where $B_0$ is the sum of energies of $2$ Hydrogenic atoms and $B_1$ is the so-called electronic interaction energy.
They can be calculated analytically. In atomic units, the first three coefficients in (\ref{1overZ}) are given by
\begin{equation}
\label{1overZ-2}
  B_0\ =\ 1 \ ,\qquad B_1\ =\ \frac{5}{8}\ = \ 0.625\ , \qquad B_2\ =\,  -0.15766642946915\ ,
\end{equation}
$B_{0,1}$ being rational numbers. Remarkably, it was proved that the expansion (\ref{1overZ}) has a finite radius of convergence \cite{Kato:1980}.

Let us take the optimal parameters $\alpha=0.929044,\,\beta=-0.254746$ obtained at $Z=2$, they remain almost constant in the interval $2\leq Z\leq10$. Expanding the function $E_{\rm var}^{(3D)}$ (\ref{Evar2}) in powers of $1/Z$ we obtain
\begin{equation}
\label{1overZVar}
  E_{\rm var}^{(3D)}(Z)\ =\ -0.994965\, Z^2 \ + \  0.603247\, Z\  - \ 0.11955 \ + \  O\bigg(\frac{1}{Z}\bigg)\ ,
\end{equation}
hence, there is a good agreement with the exact values (\ref{1overZ-2}).

Secondly, at small nuclear charges $Z$, Stillinger \cite{Stillinger:1966} proposed a Puiseux expansion for the ground state energy in the form
\begin{equation}
\label{PuiseuxGen}
\begin{split}
 E(Z)\  = \ & E_{B}\ + \ p_1 \left( Z-Z_{B} \right)
\  + \ q_{{3}} \left( Z- Z_{B}\right)^{3/2} + p_{{2}} \left( Z - {\it Z_B} \right)^{2}
 \ + \ q_{{5}} \left( Z- Z_{B} \right) ^{5/2}
\\&
\ + \ p_{{3}} \left( Z- Z_{B} \right)^{3}+q_{{7}} \left( Z- Z_{B} \right)^{7/2}
\ + \ p_{{4}} \left( Z- Z_{B} \right)^{4} + \ldots \ ,
\end{split}
\end{equation}
where $Z_B > 0$ is a certain critical charge and $E_{B}=E(Z_{B})$. This was confirmed quantitatively in \cite{TG:2011}-\cite{TLO:2016}. Moreover, the expansion (\ref{PuiseuxGen}) was derived numerically in \cite{HOAT} using highly accurate values of ground state energy in close vicinity of $Z > Z_B$ obtained variationally. It turns out that $Z_B$ is not equal to the (first) critical charge $Z_c$ where the one-electron ionization energy vanishes and, unlike the aforementioned $1/Z$ expansion, there is no a rigours proof of the convergence in the Puiseux expansion (\ref{PuiseuxGen}). Explicitly, the first coefficients in (\ref{PuiseuxGen}) are \cite{HOAT}
\[
 Z_B^{}\ =\ {0.9048539992}\ ,\qquad E_B^{}\ =\ {-0.407932489} \ , \qquad
 p_1^{}\ =\, {-1.123475} \ ,
 \]

\begin{equation}
\label{k2par}
 \ q_3^{}\ =\, -0.197785\ ,\qquad p_2^{}\ =\, -0.752842\ ,
\end{equation}
cf. \cite{TLO:2016}.

Now, we take the same optimal parameters $\alpha=0.929044,\,\beta=-0.254746$ obtained at $Z=2$, and expanding the function $E_{\rm var}^{(3D)}$ around $Z=Z_B$ we arrive to the expression
\begin{equation}
\label{PuisVar}
  E_{\rm var}^{(3D)}(Z)\ =\ - \ 0.380745 \ - \ 1.20635\,(Z-Z_B)\ - \ 0.984326\,{(Z-Z_B)}^2  \  + \  \ldots \ ,
\end{equation}
thus, in spite of the fact that (\ref{PuisVar}) does not contain terms with fractional degrees it does provide reasonable values for the first coefficients in front of the terms with integer degree.

\bigskip

\subsection{Case $d=2$}

Now, we treat the less studied planar case $d=2$. Such a system occurs in condensed matter physics. For the ground state we use the Hylleraas-type trial function $\psi_{\small \rm Hs}$ (\ref{psi1}) again.

\vspace{0.2cm}

The weight function $w$ in (\ref{Evar}) becomes
\[
w(d=2)\ \propto \ \frac{(\eta +\sigma ) (2\, \eta +\tau ) (2 \,\sigma +\tau )}{\sqrt{\eta \, \sigma\,  \tau\,  (2 \,\eta +2\, \sigma +\tau )}} \ = \ \frac{u_1\, [{(u_1 +u_3)}^2\,-\,(u_1^2-4\,u_2)]}{\sqrt{u_2\,  u_3\,  (2 \,u_1 +u_3 )}} \ .
\]
Using the $u$-variables (\ref{uvar}) we were able to calculate the expectation value $\langle r_{12}^{-1}\rangle_{\small  \psi_{\small \rm Hs}}$ in a closed-analytical form. The variational energy can be written as follows

\begin{equation}\label{Evar22D}
E_{\rm var}^{(2D)} \ = \  (\alpha ^2 \,Z^2\,-\,\beta ^2)\,\frac{A(\alpha\,Z,\,\beta,\,Z)}{B(\alpha\,Z,\,\beta,\,Z)}  \ ,
\end{equation}
where
\begin{equation}\label{}
\begin{aligned}
&  A \ =  \  -2 \beta ^5 (\beta +2)+8 (\alpha -4) \alpha ^5 Z^6+\alpha ^3 \beta  Z^4 (-9 \alpha  \beta -10 \alpha +16 \beta )+\alpha  \beta ^3 Z^2 (3 \alpha  \beta +14 \alpha +16 \beta )
\\ &
 \qquad \ - \ 3\, \alpha ^3 Z^4 (\alpha  (3 \beta -2)-16 \beta ) \sqrt{\alpha ^2 Z^2-\beta ^2} \cos ^{-1}\left(\frac{\beta }{\alpha  Z}\right) \ ,
\\ &
   B\ =  \ 2 \,\beta ^6+8\, \alpha ^6 \,Z^6+\alpha ^4 \,\beta ^2 \,Z^4-11\, \alpha ^2\, \beta ^4\, Z^2-15\, \alpha ^4\, \beta  Z^4 \sqrt{\alpha ^2 Z^2-\beta ^2} \cos ^{-1}\left(\frac{\beta }{\alpha  Z}\right) \ ,
\end{aligned}
\end{equation}

(c.f. (\ref{Evar2})). Therefore, unlike the previous 3D case, it is not an algebraic function of the parameters $\alpha$ and $\beta$. To the best of our knowledge the analytical expression (\ref{Evar22D}) is presented for the first time. At $\beta=0$, (\ref{Evar22D}) reduces to
\begin{equation}\label{E2Dal}
E_{\rm var}^{(2D)}(\beta=0) \ = \  (\alpha -4)\, \alpha \, Z^2 \ + \ \frac{3\, \pi\,  \alpha \, Z}{8}\ .
\end{equation}
(c.f. (\ref{E3Dal}),(\ref{Majorana})). We call (\ref{Evar22D}) the 2D-generalized Majorana solution. In particular, for the ground state of the Helium atom ($Z=2$) we obtain the energy
\[
E_{0}^{(2D)}(Z=2) \ = \ -11.8350\,a.u. \ ,
\]
at $\alpha=1.87638,\,\beta=-0.53717$. Hence, a nuclear-electron cusp $\nu_1=-3.75277$ and an electron-electron cusp $\nu_2=0.5372$ . It has to be compared with the \emph{exact} numerical result \cite{Hilico}:
\[
E_{\rm exact}^{(2D)}(Z=2)  \ = \ -11.8998\,a.u.  \ .
\]
It implies a small relative error ($\sim 0.5\%$) in the variational energy. Now, the denominator (\ref{Lambd}) takes the form
\begin{equation}\label{}
 \Lambda^{(2D)}(\alpha,\,\beta) \ = \ \frac{\pi  \left(\sqrt{(\alpha  Z-\beta ) (\beta +\alpha  Z)} \left(-2 \beta ^4+8 \alpha ^4 Z^4+9 \alpha ^2 \beta ^2 Z^2\right)-15 \alpha ^4 \beta  Z^4 \sec ^{-1}\left(\frac{\alpha  Z}{\beta }\right)\right)}{16 \alpha ^2 Z^2 ((\alpha  Z-\beta ) (\beta +\alpha  Z))^{7/2}} \ .
\end{equation}

By taking the polynomial function $P(r_1,\,r_2,\,r_{12})$ as the inter-electronic distance $P=r_{12}$ and using the relation (\ref{relat}) we obtain
{\small
\begin{equation}\label{}
  \langle \,r_{12}\, \rangle \, = \, \frac{-4 \beta ^7-81 \alpha ^6 \beta  Z^6+53 \alpha ^4 \beta ^3 Z^4+32 \alpha ^2 \beta ^5 Z^2+15 \alpha ^4 Z^4 \left(6 \beta ^2+\alpha ^2 Z^2\right) \sqrt{\alpha ^2 Z^2-\beta ^2} \sec ^{-1}\left(\frac{\alpha  Z}{\beta }\right)}{2 (\alpha^2  Z^2-\beta^2 )\left(2 \beta ^6+8 \alpha ^6 Z^6+\alpha ^4 \beta ^2 Z^4-11 \alpha ^2 \beta ^4 Z^2-15 \alpha ^4 \beta  Z^4 \sqrt{\alpha ^2 Z^2-\beta ^2} \sec ^{-1}\left(\frac{\alpha  Z}{\beta }\right)\right)}\ .
\end{equation}
}
\subsubsection{$\frac{1}{Z}$ Expansion}

In this subsection, it is worth analyzing the 2D GMS with respect to the $1/Z$ expansion. Assuming $Z \rightarrow \infty$, we immediately calculate the ground state energy in perturbation theory in the small parameter $1/Z$,
\begin{equation}
\label{1overZ2D}
  E^{(2D)}(Z)\ =\ -C_0\, Z^2 \ + \  C_1\, Z\  + \ C_2 \ + \  O\bigg(\frac{1}{Z}\bigg)\ ,
\end{equation}
where $C_0$ is the sum of energies of $2$ two-dimensional Hydrogenic atoms and $C_1$ is the so-called electronic interaction energy. They can be calculated analytically,
\begin{equation}
\label{1overZ-22D}
  C_0\ =\ 4 \ ,\qquad C_1\ =\ \frac{3\, \pi }{4 }\ \approx \ 2.3562\ ,
\end{equation}

c.f. (\ref{1overZ}), (\ref{1overZ-2}). Now, let us take the optimal parameters $\alpha=1.876383,\,\beta=-0.537171$ obtained at $Z=2$ (they remain almost constant in the interval $2\leq Z\leq10$, see Fig. \ref{varp}). Expanding the GMS $E_{\rm var}^{(2D)}$ (\ref{Evar22D}) in powers of $1/Z$ we obtain
\begin{equation}
\label{1overZVar2D}
  E_{\rm var}^{(2D)}(Z)\ =\ -3.98472\, Z^2 \ + \  2.28879\, Z\  - \ 0.490893 \ + \ O\bigg(\frac{1}{Z}\bigg)\ ,
\end{equation}
c.f. (\ref{1overZVar}). Not surprisingly, a good agreement with the exact values occurs.

\subsection{Case $d=4$}

Now, we move to the case $d=4$. The variational energy with respect to (\ref{psi1}) is given by

\begin{equation}\label{Evar24D}
E_{\rm var}^{(4D)} \ = \  (\alpha ^2 \,Z^2\,-\,\beta ^2)\,\frac{{\cal A}(\alpha\,Z,\,\beta,\,Z)}{{\cal B}(\alpha\,Z,\,\beta,\,Z)}  \ ,
\end{equation}
where
\begin{equation}\label{}
\begin{aligned}
&  {\cal A} \ =  \  2 q_1 [16 \beta ^7 (3 \beta +2)+256 \alpha ^7 (4-3 \alpha ) Z^8+13 \alpha ^5 \beta  Z^6 (-303 \alpha  \beta +102 \alpha +608 \beta )
\\ &
\qquad \ + \ 10 \alpha ^3 \beta ^3 Z^4 (15 \alpha  \beta +74 \alpha +128 \beta )-8 \alpha  \beta ^5 Z^2 (27 \alpha  \beta +26 \alpha +16 \beta )]
\\ &
+105 \alpha ^4 Z^5 \left(8 \beta ^2 (3 \alpha  \beta -2 \alpha -8 \beta )+\alpha ^2 Z^2 (21 \alpha  \beta -2 \alpha -32 \beta )\right) \left(\pi  \alpha ^2 Z^2-2 \csc ^{-1}\left(\frac{\alpha  Z}{\beta }\right)\right)  \ ,
\\ &
   {\cal B}\ =  \  945 \alpha ^5 \beta  Z^5 \left(8 \beta ^2+3 \alpha ^2 Z^2\right) \left(\pi  \alpha ^2 Z^2-2 \csc ^{-1}\left(\frac{\alpha  Z}{\beta }\right)\right)
\\ &
\qquad \ -6 \,q_1\, \left(16 \beta ^8+256 \alpha ^8 Z^8+2639 \alpha ^6 \beta ^2 Z^6+690 \alpha ^4 \beta ^4 Z^4-136 \alpha ^2 \beta ^6 Z^2\right) \ ,
\end{aligned}
\end{equation}

here $q_1=\sqrt{1-\frac{\beta ^2}{\alpha ^2 Z^2}}$. Hence, like for $d=2$, it is not an algebraic function of the variational parameters. As mentioned in the Introduction, to evaluate the energy functional for even $d$ we employed the $u$-variables (\ref{uvar}) instead of the perimetric ones. At $\beta=0$, the analytical expression (\ref{Evar24D}) reduces to
\begin{equation}\label{}
E_{\rm var}^{(4D)}(\beta=0) \ = \ 1 \ - \ \frac{4}{3 \,\alpha }\ + \ \frac{35\, \pi\,  \alpha \, Z}{256} \ .
\end{equation}
(c.f. (\ref{Majorana}),(\ref{E3Dal}),(\ref{E2Dal})). Thus, we call (\ref{Evar24D}) the 4D-generalized Majorana solution. Accordingly, for the ground state of the Helium atom ($Z=2$) we arrive to the value
\[
E_{0}^{(4D)}(Z=2)  \ = \ -1.26809\,a.u. \ ,
\]
at $\alpha=0.6160175,\,\beta=-0.1650859$. Hence, a nuclear-electron cusp $\nu_1=-1.232035$ and an electron-electron cusp $\nu_2=0.1651$ . It has to be compared with the \emph{exact} numerical result \cite{Duan}:
\[
E_{\rm exact}^{(4D)}(Z=2)  \ = \ -1.27364\,a.u.  \ .
\]

It implies a small relative error ($\sim 0.43\%$) in the variational energy.

\subsubsection{$\frac{1}{Z}$ Expansion}

Assuming $Z \rightarrow \infty$, let us determine the ground state energy in perturbation theory in the small parameter $1/Z$,
\begin{equation}
\label{1overZ4D}
  E^{(4D)}(Z)\ =\ -D_0\, Z^2 \ + \  D_1\, Z\  + \ D_2 \ + \  O\bigg(\frac{1}{Z}\bigg)\ ,
\end{equation}
where $D_0$ is the sum of energies of $2$ four-dimensional Hydrogenic atoms and $D_1$ is the so-called electronic interaction energy. They also can be found exactly,
\begin{equation}
\label{1overZ-24D}
  D_0\ =\ \frac{4}{9} \ \approx \ 0.444 \ ,\qquad D_1\ =\ \frac{35\, \pi}{384}\ \approx \ 0.28634\ ,
\end{equation}
c.f. (\ref{1overZ-22D}), (\ref{1overZ-2}). Taking the optimal parameters $\alpha=0.6160175,\,\beta=-0.1650859$ obtained at $Z=2$ and expanding for large $Z\rightarrow \infty$ the GMS $E_{\rm var}^{(2D)}$ (\ref{Evar24D}) in powers of $1/Z$, we arrive to the expression
\begin{equation}
\label{1overZVar4D}
  E_{\rm var}^{(4D)}(Z)\ =\ -0.441879\, Z^2 \ + \  0.275362\, Z\  - \ 0.0525401 \ + \ O\bigg(\frac{1}{Z}\bigg)\ ,
\end{equation}
thus, in excellent agreement with the exact values.

\subsection{Case $d=5$}

For the case $d=5$, the weight function $w$ appearing in $E_{\rm var}$ (\ref{Evar}) becomes $w(d=5) \propto \eta \, \sigma\,  \tau\,  (\eta +\sigma ) (2\, \eta +\tau ) (2 \, \sigma +\tau ) (2 \,\eta +2 \sigma +\tau )$. The variational energy with respect to (\ref{psi1}) reads

\begin{equation}\label{Evar25D}
E_{\rm var}^{(5D)} \ = \  (\, \beta \, +\, \alpha  \,Z)\,\frac{\tilde A(\alpha\,Z,\,\beta,\,Z)}{\tilde B(\alpha\,Z,\,\beta,\,Z)}  \ ,
\end{equation}
where
\begin{equation}\label{}
\begin{aligned}
& \tilde A \ =  \ \alpha ^3\, Z^3\, (\,64 (\alpha -1) Z\,+\,21) \ + \ \alpha ^2\, \beta\,  Z^2 \,(14 (5\, \alpha \,-\,2) Z \,+\,19)
\\ &
\qquad \ + \ \beta ^3\, (14 \,\alpha \, Z\,+\,1)\ + \ \alpha\,  \beta ^2 \,Z\, ((42\, \alpha -4) Z\,+\,7) \ + \ 2\, \beta ^4 \ ,
\\ &
\tilde B \ =  2\, \left(\,32\, \alpha ^3\, Z^3\ + \ 25\, \alpha ^2 \,\beta\,  Z^2 \ + \ 8\, \alpha  \,\beta ^2\, Z \ +\ \beta ^3\,\right) \ \ ,
\end{aligned}
\end{equation}

(c.f. (\ref{Evar2})) thus, it is a polynomial function of the variational parameters. At $\beta=0$, the energy (\ref{Evar25D}) reduces to
\begin{equation}\label{}
E_{\rm var}^{(5D)}(\beta=0) \ = \  \frac{1}{64} \alpha \, Z^2 \,\left(\,64\, \alpha\,  Z^2\ - \ 64\, Z \ + \ 21\,\right) \ .
\end{equation}
(c.f. (\ref{Majorana}),(\ref{E3Dal})). The expression (\ref{Evar25D}) can be called the 5D-GMS.

Eventually, for the Helium atom ($Z=2$) we obtain the ground state energy
\[
E_{0}^{(5D)}(Z=2)  \ = \ -0.7077\,a.u.
\]
at $\alpha=0.460444,\,\beta=-0.121678$. Hence, a nuclear-electron cusp $\nu_1=-0.920888$ and an electron-electron cusp $\nu_2=0.1217$ . It has to be compared with the \emph{exact} numerical result \cite{Duan}:
\[
E_{\rm exact}^{(5D)}(Z=2)  \ = \ -0.71050\,a.u.  \ .
\]
In this case, the relative error in the variational energy continues to be small $\sim 0.43\%$.

\vspace{0.15cm}

\begin{center}
\begin{figure}[h]
\includegraphics[scale=0.28]{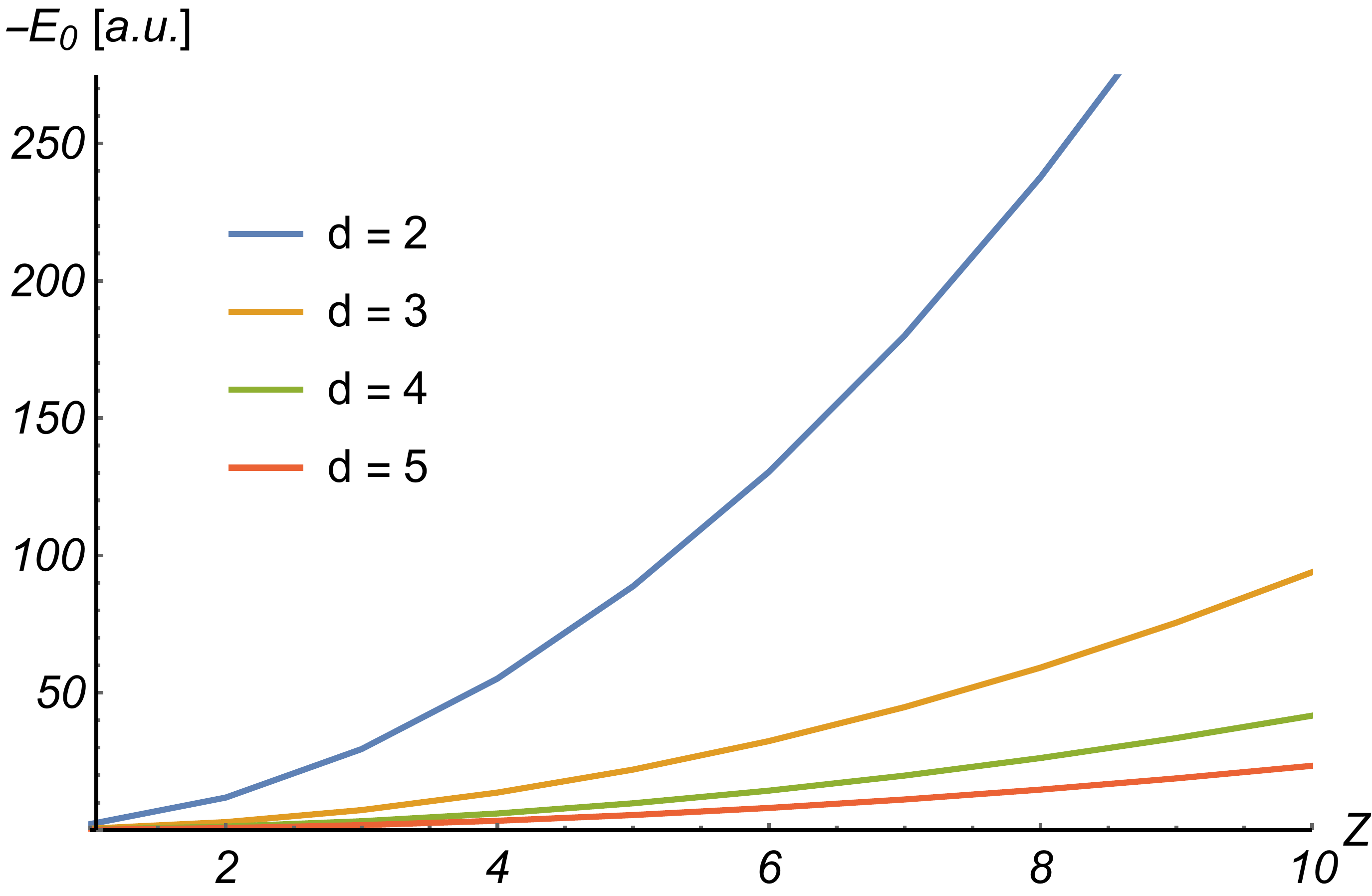}  \hspace{0.2cm} \includegraphics[scale=0.28]{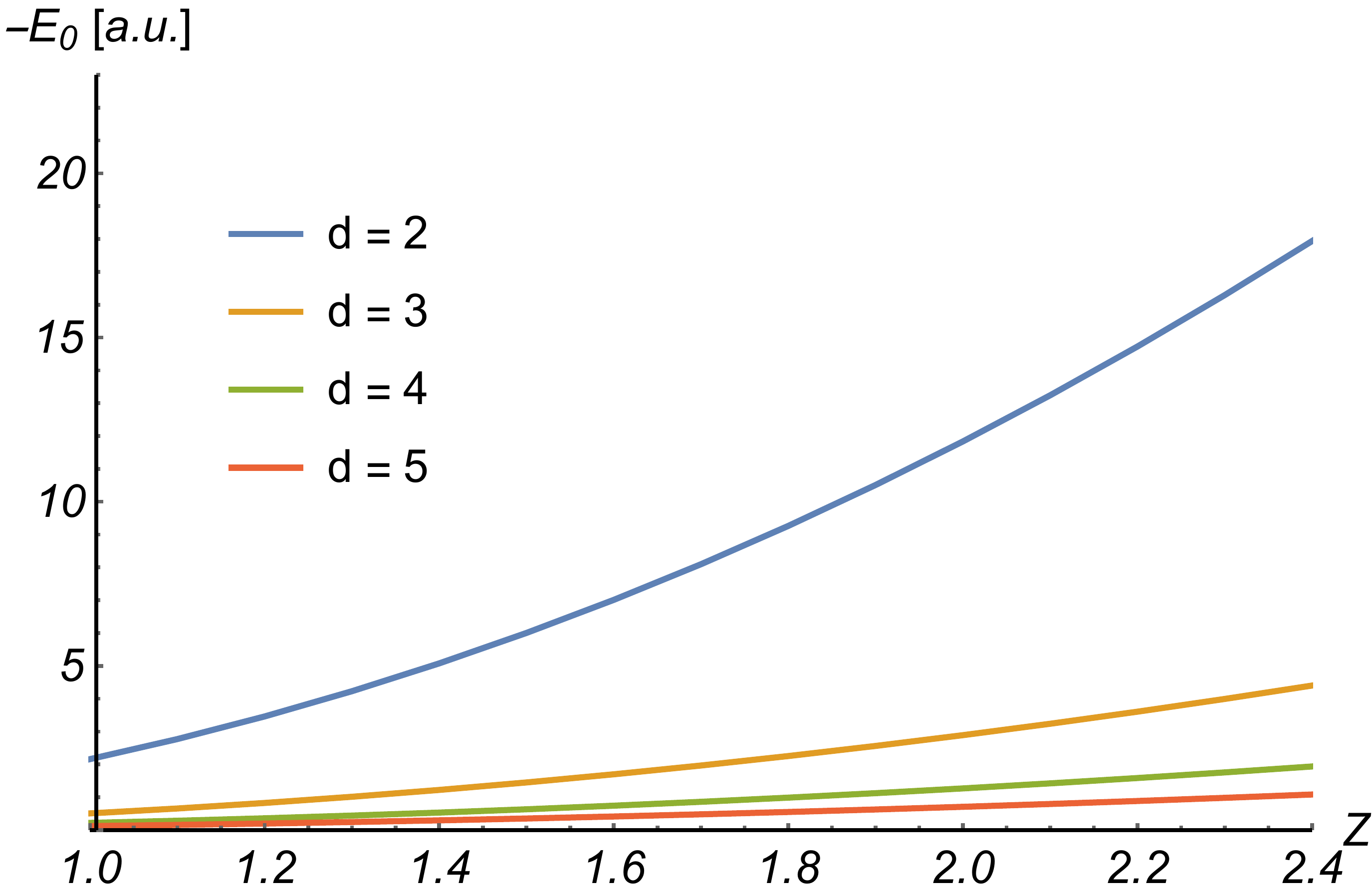}
\caption{Ground state energy: $-E_0$ vs. $Z$ for $1\leq Z \leq 10$ in static approximation (infinite nuclear mass case). The variational energy $E_0$ is calculated with the 2-parametric Hylleraas-type function (\ref{psi1}). The energy is in atomic units (a.u.), (color online).}
\label{enerp}
\end{figure}
\end{center}

\begin{center}
\begin{figure}[h]
\includegraphics[scale=0.28]{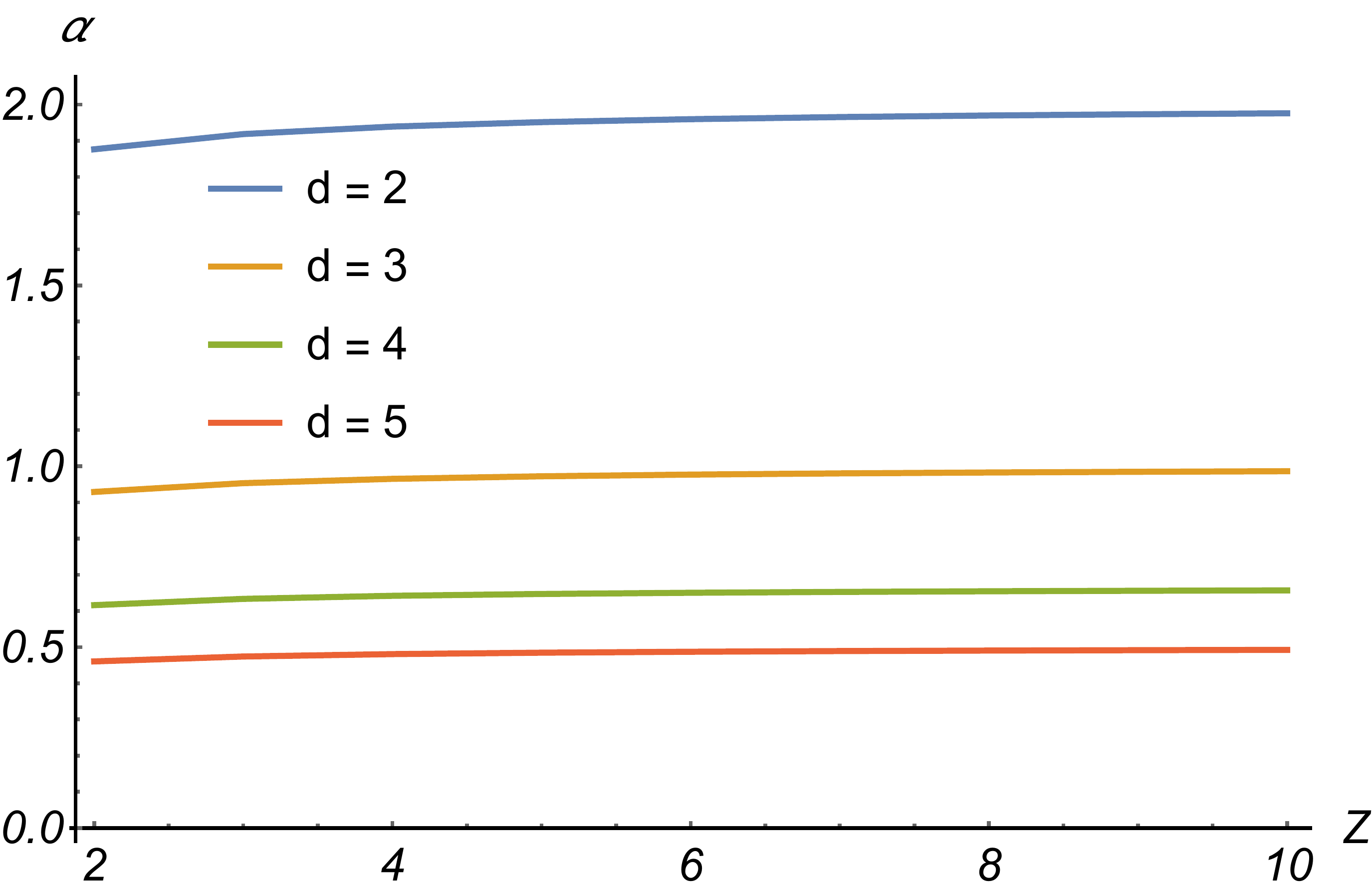}  \hspace{0.2cm} \includegraphics[scale=0.28]{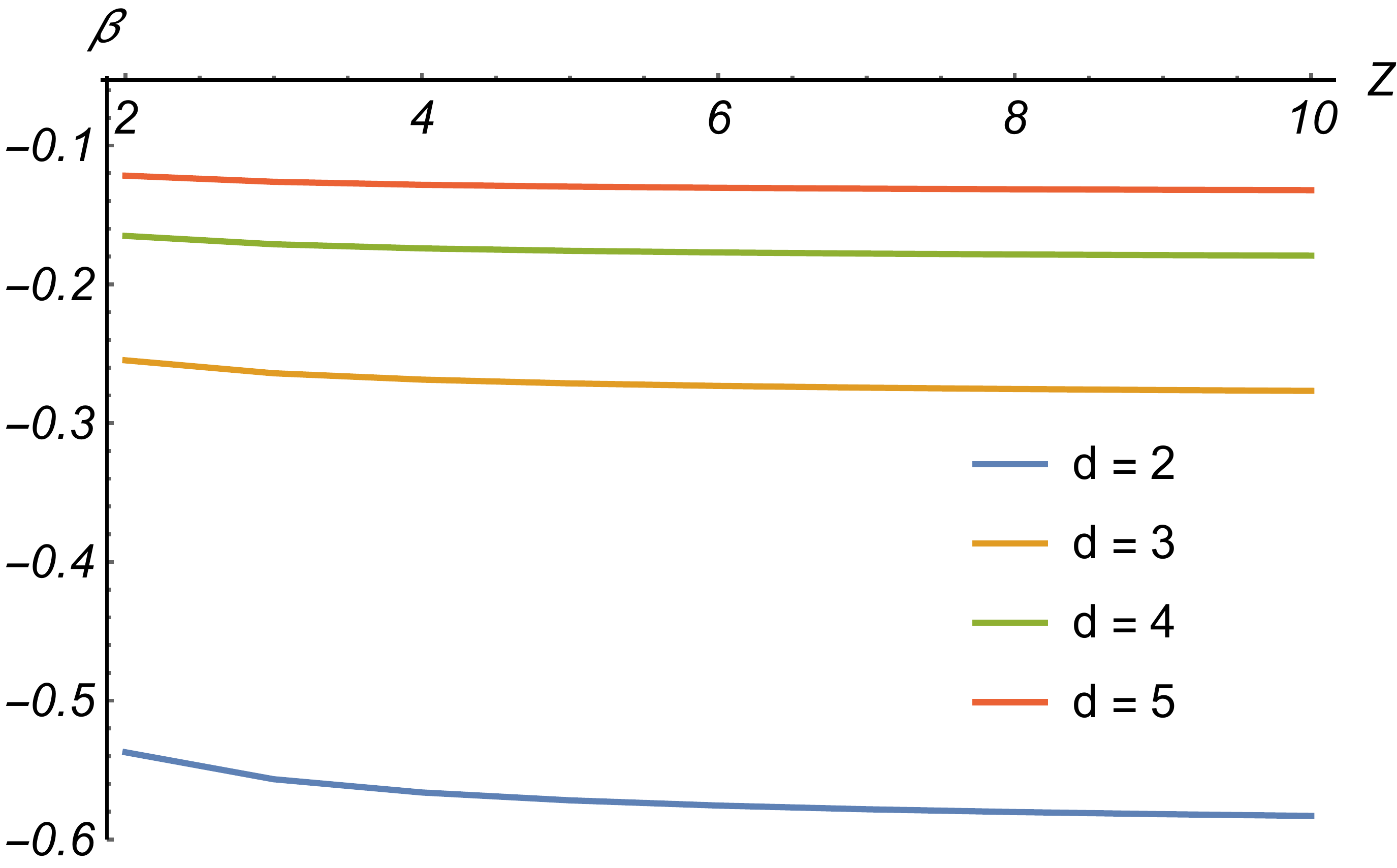}
\caption{Ground state energy: non-linear parameters $\alpha$ vs $Z$ (left) and $\beta$ vs $Z$ (right) for $2\leq Z \leq 10$ in static approximation (infinite nuclear mass case). These parameters define the 2-parametric Hylleraas-type trial function (\ref{psi1}), (color online). }
\label{varp}
\end{figure}
\end{center}

In Table \ref{Tab1}, for the lowest values of $d=2,3,4,5$ we display the ground state energy $E_0=E_0(Z)$ for a $d$-dimensional helium-like ion in the infinite nuclear mass approximation. The results were obtained using the 2-parametric Hylleraas-type trial function (\ref{psi1}). Due to the analyticity of the formalism, one can treat $Z$ as a continuous variable in the non-relativistic domain $2\leq Z \leq 10$. The corresponding plots are shown in Figure \ref{enerp}. The associated variational parameters $\alpha(Z)$, $\beta(Z)$ are presented in Figure \ref{varp}. They are smooth monotonic functions of the nuclear charge $Z$.

\clearpage

\begin{table}[h]
\caption{The numerical calculation for the ground state energy $E_0$ in a $d$-dimensional helium-like ion in the infinite nuclear mass approximation. The expectation value $\langle r_{12} \rangle$ also calculated from the 2-parametric Hylleraas-type trial function (\ref{psi1}). Numerical results presented in atomic units.}
\centering
\begin{tabular}{|c||c|c|c|c||c|c|c|c|}
\hline
&  \multicolumn{4}{c|}{ $E_0\, (a.u.)$} & \multicolumn{4}{c|}{ $\langle r_{12} \rangle \, (a.u.)$ } \\
\hline
\hline
\hspace{0.15cm} $Z$ \hspace{0.15cm} & \hspace{0.3cm}  $d=2$ \hspace{0.3cm} & \hspace{0.3cm} $d=3$ \hspace{0.3cm} & \hspace{0.3cm} $d=4$ \hspace{0.3cm} & \hspace{0.3cm} $d=5$ \hspace{0.3cm} & \hspace{0.3cm} $d=2$ \hspace{0.3cm} & \hspace{0.3cm} $d=3$ \hspace{0.3cm} & \hspace{0.3cm} $d=4$ \hspace{0.3cm} & \hspace{0.3cm} $d=5$ \hspace{0.3cm}  \\
\hline
2 & -11.8350 & -2.88961 & -1.26809 & -0.707704 & 0.4668 & 1.385 & 2.754 & 4.573  \\
\hline
3 & -29.4889 & -7.26681 & -3.20486 & -1.79411  & 0.2869 & 0.8515 & 1.692 & 2.808  \\
\hline
4 & -55.1377 & -13.6429 & -6.03008 & -3.38029  & 0.2067 & 0.6137 & 1.219 & 2.024 \\
\hline
5 & -88.7845 & -22.0185 & -9.74400 & -5.46636  & 0.1615 & 0.4795 & 0.9530 & 1.581  \\
\hline
6 & -130.430 & -32.3939 & -14.3467 & -8.05240  & 0.1325 & 0.3934 & 0.7819 & 1.297  \\
\hline
7 & -180.075 & -44.7692 & -19.8382 & -11.1384  & 0.1123 & 0.3335 & 0.6629 & 1.100  \\
\hline
8 & -237.720 & -59.1445 & -26.2187 & -14.7243  & 0.0974 & 0.2894 & 0.5753 & 0.9548  \\
\hline
9 & -303.365 & -75.5197 & -33.4879 & -18.8103  & 0.0860 & 0.2556 & 0.5081 & 0.8433  \\
\hline
10& -377.009 & -93.8948 & -41.6461 & -23.3963 & 0.0770 & 0.2289 & 0.4549 & 0.7551  \\
\hline
\end{tabular}
\label{Tab1}
\end{table}

\section{The Shannon Entropy and the (first) critical charge $Z_c$}

So far we have discussed the ability of the GMS to
account for reasonably accurate values of the ground-state energy of He-like
systems with clamped nuclei in $d= 2,3,4$ and $5$ dimensions. In this context, it is
worth analyzing the behavior of another two relevant properties derived from the
electronic density and nuclear charge, namely the Shannon entropy and the
(first) critical charge $Z$.

\subsection{The Shannon Entropy}

In many-electron systems the Shannon entropy \cite{Shannon} has been used as a measure of the uncertainty in the
position of one electron, which is related with the degree of
localization-delocalization of the particle. When the Shannon entropy is
small the uncertainty in the position of the particle is also small, and the
particle is more localized and \emph{vice-versa}, see \cite{Esta}-\cite{Aquino3} an references therein. In the present study the
Shannon entropy will be used as a measure of the localization (delocalization) of the
particle.

More precisely, we are to determine the Shannon entropy within the formalism developed in
previous sections. For the ground state of a $d$-dimensional Helium-like ion we shall compute the Shannon entropy in the Hylleraas coordinates $r_{1},r_{2}$ and $r_{12}$ using the optimal trial function (\ref{psi1}).

A $d$-dimensional two electron system is completely described by a wave
function of the form $\Psi ({\bf r}_{1},{\bf r}_{2},\sigma _{1},\sigma _{2})$%
, where ${\bf r}=(x_{1},x_{2},\ldots ,x_{d})$. The main quantity involved in the
Shannon entropy is the single-particle density $\rho ^{(d)}({\bf r})$,
defined as:

\begin{equation}\label{spd}
\rho ^{(d)}({\bf r}_{1})=\sum_{\sigma _{i}=-1/2}^{+1/2}\int \left\vert \Psi (%
{\bf r}_{1},{\bf r}_{2},\sigma _{1},\sigma _{2})\right\vert ^{2}\times d{%
\bf r}_{2} \ ,
\end{equation}

where $\sigma _{i}$ represents the spin coordinate of the i$th$ particle.
The wave function is normalized and it is antisymetric under the exchange of
the particles. The single particle density \ $\rho ^{(d)}({\bf r})$ is also
normalized to the unity.

Since the ground state function of a Helium-like ion solely depends on the Hylleraas coordinates, the single-particle density (\ref{spd}) can be written in the form

\begin{equation}
\rho^{(d)} (r_1)\ = \ \frac{1}{s_{d-1}(r_1)}\int_{0}^{\infty}\int_{|r_1-r_2|}^{r_1+r_2}\,|\Psi (r_1,r_2,r_{12})|^2 \, dR\ ,
\label{rho1S}
\end{equation}

where $s_{d-1}(r)=\omega _{d}r^{d-1}$, $\omega _{d}=2\pi ^{d/2}/\Gamma (d/2)$
is the area of the surface of a unit sphere in $d$-dimensions, and

\begin{equation*}
dR \ = \ \frac{2^d \,\pi^{d-1}}{(d-2)!} \,r_2\, r_{12} \,S^{d-3}\, dr_2 \,dr_{12}\ .
\end{equation*}

In (\ref{rho1S}), the function $\Psi (r_1,r_2,r_{12})$ is taken from (\ref{psi1}). Accordingly, the Shannon entropy is given by

\begin{equation}
S_{r}^{(d)}\ = \ -\int_{0}^{\infty }\rho ^{(d)}(r)\,\ln \rho ^{(d)}(r)\,\omega _{d}\,r^{d-1}\,dr \ .
\end{equation}

In three dimensional space $(d=3)$ the single particle density and the Shannon entropy reduce to
the well-known expressions

\begin{equation}
\rho(r_1) \ = \ \frac{2\,\pi}{r_1}\int_{0}^{\infty}\int_{|r_1-r_2|}^{r_1+r_2}\,|\Psi(r_1,r_2,r_{12})|^2\, r_2\, r_{12}\, dr_2\, dr_{12}\ ,
\end{equation}

and

\begin{equation}
S_{r}\ = \ -4\,\pi\,\int_0^{\infty}\,\rho(r)\,\ln{\rho(r)}\, r^2\, dr\ .
\end{equation}

\begin{figure}[h]
\includegraphics[scale=1.1]{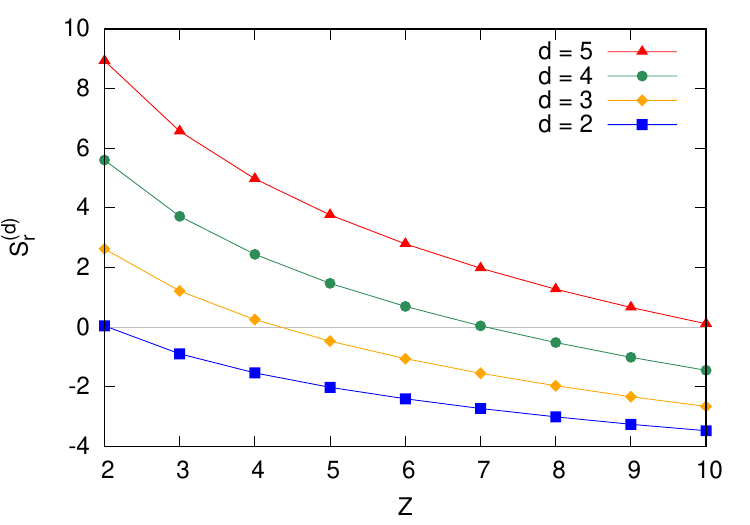}
\caption{Shannon entropy $S_{r}^{(d)}$ for two-electron systems as a function of the nuclear charge $Z$ for
$d= 2, 3 ,4$ and $5$ dimensions. Numerical values marked by bullets, the lines are a guide to the eye.}
\label{fshan}
\end{figure}

Table~\ref{Tshan} displays the Shannon entropy as a function of the nuclear charge
$Z=2, \dots , 10$  for $d=2, 3, 4$ and $5$ (see also Figure~\ref{fshan}).  Two important
features appear in the behavior of the Shannon entropy: $i$) for each fixed value of $d$,  $S^{(d)}_{r}$
is a strictly decreasing function of $Z$  and, $ii$) in agreement with the results for the $d$-dimensional
hydrogen atom~\cite{YAD:2020},  $S^{(d)}_{r}$ is a strictly increasing function of $d$.
Column 4 of Table~\ref{Tshan} shows the results of the Shannon entropy for $d=3$  calculated in
\cite{LH:2015} (with 444 basis functions, label $a$) and~\cite{NA:2020} (with a three-parameter wave function, label $b$). For
$Z=2,3$ the relative error is $\sim 3\%$ and for $Z=4, \dots,10$ the relative error decreases from
$\sim10\%$ to $\sim 0.4\%$

{\small
\begin{table}[h]
\begin{center}
\begin{tabular}{|c|c|c|c|c|}
\hline\hline
\hspace{0.2cm} $Z\quad$ & \hspace{0.2cm} $d=2$ \hspace{0.2cm} & \hspace{0.3cm} $d=3$ \hspace{0.3cm} & \hspace{0.2cm} $d=4$ \hspace{0.2cm} & \hspace{0.2cm} $d=5$  \hspace{0.2cm} \\
\hline\hline
2 &  0.0425&  2.6194\quad  2.7051$^{\, a}$& 5.5974& 8.9218\\
3 & -0.8952&  1.2090\quad  1.2553$^{\, a}$& 3.7135& 6.5642\\
4 & -1.5318&  0.2524\quad  0.2824$^{\, b}$& 2.4368& 4.9673\\
5 & -2.0142& -0.4719\quad -0.4489$^{\, b}$& 1.4703& 3.7585\\
6 & -2.4027& -1.0551\quad -1.0364$^{\, b}$& 0.6923& 2.7857\\
7 & -2.7278& -1.5432\quad -1.5274$^{\, b}$& 0.0413& 1.9717\\
8 & -3.0074& -1.9629\quad -1.9493$^{\, b}$&-0.5185& 1.2717\\
9 & -3.2527& -2.3310\quad -2.3190$^{\, b}$&-1.0095& 0.6579\\
10& -3.4712& -2.6588\quad -2.6482$^{\, b}$&-1.4468& 0.1112\\
\hline\hline
\end{tabular}
\end{center}
\caption{Shannon entropy $S_{r}^{(d)}$ for two-electron systems with nuclear charge $Z$ in a
$d$-dimensional space.}
\label{Tshan}
\end{table}
}

The Shannon entropy $S_r^{(d)}(Z)$ for $d=3, 4$ and $5$ (Table~\ref{Tshan})
can be obtained from the results for $d=2$, namely $S_r^{(2)}(Z)$, with the empirical interpolation
\begin{equation}
S_r^{(d)}(Z)\ = \ S_0^{(d)} \ + \ \frac{d}{2}S_r^{(2)}(Z)\ + \ \frac{(d-1)(d-2)+2}{200\, Z}, \hspace{1cm} d=3,4,5,
\label{sinter}
\end{equation}
where $S_0^{(3)}=2.5455$, $S_0^{(4)}=5.4919$, $S_0^{(5)}=8.7805$ and $Z$ being the nuclear charge.
The absolute error in the Shannon entropy calculated with (\ref{sinter}) is $\lesssim 10^{-3}$ compared to the numerical results of Table~\ref{Tshan}.

\clearpage

\subsection{The (first) critical charge $Z_c$}
\label{secCrc}

Considering the nuclear charge as a continuous parameter in the Helium-like sequence, an important characteristic of the system is the so call (first) critical charge $Z_c$ above of which, $Z > Z_c$, the system gets bound. Explicitly, the charge $Z_c$ is defined by the relation

\begin{equation}\label{}
E_{\rm var}^{(d)}(Z_c) \ + \  \frac{1}{2} \frac{Z_c^2}{{(1 \,+\, [d-3]/2)}^2} \ = \ 0 \ ,
\end{equation}

where $-\frac{1}{2} \frac{Z_c^2}{{(1 \,+\, [d-3]/2)}^2}$ corresponds to ground state energy of the Hydrogen atom in $d-$dimensions.

The aforementioned generalized Majorana formulas (\ref{Evar2}), (\ref{Evar22D}), (\ref{Evar24D}) and (\ref{Evar25D}) give a rather accurate value of $Z_c$. Our variational calculations using (\ref{psi1}) provide the following values

\begin{equation}\label{}
Z_c^{(d=2)} \ \approx \ 0.8931 \ ; \,\quad Z_c^{(d=3)} \ \approx \ 0.9791 \ ; \,\quad  Z_c^{(d=4)} \ \approx \ 1.023 \ ; \,\quad  Z_c^{(d=5)} \ \approx \ 1.049 \ .
\end{equation}

In particular, at $d=3$ the relative error with respect to the exact numerical result\cite{Estienne}-\cite{HOAT}
\[
\quad Z_{c,{\rm exact}}^{(d=3)} \ = \ 0.91102822\ ,
\]
is $\approx 0.07\%$ . Below, we will show that the addition of a certain variational parameter in (\ref{psi1}) does not break the analyticity (the functional of energy can be evaluated exactly) and one can easily check that it makes the latter relative error $\approx 0.07\%$ decrease up to $\approx 0.016$.

\section{Three-body Coulomb system: finite mass effects}

In this Section, we will study in full generality the finite mass effects for a $d-$dimensional three-body Coulomb system of arbitrary charges. The study is restricted to the ground state of the system. To this end, we also employ the variational method with a more general correlated Hylleraas-type trial function.

In the case of 3 particles with arbitrary masses ($m_1,m_2,M$) and charges ($e_1,e_2,Z$), respectively, the generalization of the operator ${\cal H}_r$ (\ref{Hr}) is given by \cite{TME3-d}
\begin{equation}\label{Hrm}
  {\cal H}_{m,r} \ = \ -\frac{1}{2}\Delta_{m,r}^{(3)}(r_1,\,r_2,\,r_{12})  \ + \ \frac{e_1\,Z}{r_{1}} \ + \ \frac{e_2\,Z}{r_{2}}\ + \ \frac{e_1\,e_2}{r_{12}}\ ,
\end{equation}
where
\[
r_{13} \ \equiv \ r_1 \quad , \qquad r_{23} \  \equiv \  r_2 \ ,
\]
and
\begin{equation}
\begin{split}
  \Delta_{m,r}^{(3)}\  = & \   \frac{m_1+M}{m_1\,M}\frac{1}{r_{1}^{d-1}}\frac{\partial}{\partial r_{1}}\left(r_{1}^{d-1}\frac{\partial}{\partial r_{1}}\right)\ + \ \frac{m_2+M}{m_2\,M}\frac{1}{r_{2}^{d-1}}\frac{\partial}{\partial r_{2}}\left(r_{2}^{d-1}\frac{\partial}{\partial r_{2}}\right)
\\ &
   + \ \frac{m_1+m_2}{m_1\,m_2}\frac{1}{r_{12}^{d-1}}\frac{\partial}{\partial r_{12}}\left(r_{12}^{d-1}\frac{\partial}{\partial r_{12}}\right)
   + \ \frac{r_{1}^{2}+r_{12}^{2}-r_{2}^{2}}{m_1\,r_{1}\,r_{12}}\,\frac{\partial}{\partial r_{12}}\left(\frac{\partial}{\partial r_{1}}\right)
\\ &
   + \  \frac{r_{1}^{2}+r_{12}^{2}-r_{1}^{2}}{m_2\,r_{2}\,r_{12}}\,\frac{\partial}{\partial r_{12}}\left(\frac{\partial}{\partial r_{2}}\right)
   + \  \frac{r_{1}^{2}+r_{2}^{2}-r_{12}^{2}}{M\,r_{1}\,r_{2}}\,\frac{\partial}{\partial r_{1}}\left(\frac{\partial}{\partial r_{2}}\right) \ .
\end{split}
\end{equation}
Thus, the problem is still 3-dimensional and corresponds to a three-dimensional particle moving in a curved space. Putting $m_1=m_2=1$, $M\rightarrow \infty$, $e_1=e_2=-1$, the reduced Hamiltonian ${\cal H}_{m,r}$ (\ref{Hrm}) becomes the static-nuclei Hamiltonian ${\cal H}_{r}$ (\ref{Hr}).\\

The operator (\ref{Hrm}) is essentially self-adjoint with respect to the same radial measure $dV$ (\ref{rm}). Now, just as for the ground state we take the Hylleraas-type trial function

\begin{equation}\label{psi1m}
  \psi_{\rm m} \ = \ (1 \ + \ {\hat P}_{12})\,e^{-\alpha_1\,Z\,\,r_{13}\,-\,\alpha_2\,Z\,r_{23}\,-\,\beta\,r_{12} } \ ,
\end{equation}

where $\alpha_{1(2)}$ and $\beta$ are variational parameters such that $\psi_{\rm m}$ is square integrable ($\cal S$-bound states). The factor $1+{\hat P}_{12}$ in (\ref{psi1m}) accounts for the spatial symmetric combination associated with singlet states. In perimetric coordinates

\begin{equation}\label{psimm}
  \psi_{\rm m} \ = \ (1 \ + \ {\hat P}_{12})\,\exp \left(-\frac{1}{4}\, \alpha_1\,Z\, (2 \,\sigma +\tau )\,-\,\frac{1}{4} \, \alpha_2\,Z \,(2\, \eta +\tau )\,-\, \frac{1}{2}\, \beta \, (\eta +\sigma )\right) \ .
\end{equation}

It is worth mentioning that explicit analytical formulas for the expectation value $\langle Q\rangle_{\small  \psi_{\small \rm m}}$, for any multi-variable polynomial function $Q=Q(r_1,r_2,r_{12})$, can be obtained by simple differentiation of the denominator Den$E_{\rm var}[\psi_{\small \rm m}]$ in (\ref{Evar}) which is a function of $\alpha_{1(2)}$ and $\beta$ only. From (\ref{Evar}), we immediately obtain the relation

\begin{equation}\label{}
  \langle \, Q(r_1,r_2,r_{12})\, \rangle \ = \   \frac{1}{\text{Den}E_{\rm var}} Q\bigg(-\frac{1}{2\,Z}\partial_{\alpha_1},\,-\frac{1}{2\,Z}\partial_{\alpha_2},\,-\frac{1}{2}\partial_{\beta}\bigg)\,\text{Den}E_{\rm var} \ ,
\end{equation}

where $\partial_{\alpha_{1(2)}} \equiv \frac{\partial}{\partial \,\alpha_{1(2)}},\,\partial_{\beta} \equiv \frac{\partial}{\partial \,\beta}$. It implies that $E_{\rm var}$ plays the role of a generating function. The basic object appearing in the variational calculations is of the form

\begin{equation}\label{LambdM}
 {\cal I}(a,\,b,\,c) \ = \ \int_{0}^{\infty}\int_{0}^{\infty}\int_{|r_1-r_2|}^{r_1+r_2}\, e^{-a\,r_{13}\,-\,b\,r_{23}\,-\,c\,r_{12} }\,dr_{1}\,dr_{2}\,dr_{12}\ = \ \frac{16}{(a+b) (a+c) (b+c)} \ ,
\end{equation}

$a+b>0,a+c>0,b+c>0$. Hence, from (\ref{LambdM}) we arrive to the useful expression

\begin{equation}\label{}
\begin{aligned}
 \langle \, r_1^n\,r_2^m\,r_{12}^k\, \rangle \ & \equiv   \ \int_{0}^{\infty}\int_{0}^{\infty}\int_{|r_1-r_2|}^{r_1+r_2}\, e^{-a\,r_{13}\,-\,b\,r_{23}\,-\,c\,r_{12} }\,r_1^n\,r_2^m\,r_{12}^k\,dr_{1}\,dr_{2}\,dr_{12}
\\ &
  \ = \  (-1)^{n+m+k}\,\partial^n_{a}\,\partial_{b}^m\,\partial_{c}^k\,{\cal I}(a,\,b,\,c) \ ,
 \end{aligned}
\end{equation}

here $\partial^n_{x} \equiv \frac{\partial^n}{\partial \,x^n}$.

\subsection{Case $d=3$: concrete results}

Here we consider the three-dimensional ($d=3$) case in detail. The variational energy $E_{\rm m,var}^{(3D)}=E_{\rm var}^{(3D)}[\psi_{\rm m}]$ is a rational function of the three parameters $\alpha_{1(2)}$ and $\beta$. In general, it takes the form
\begin{equation}\label{Em3Dv}
  E_{\rm m,var}^{(3D)} \ = \ \frac{P_{10}(Z)}{P_{8}(Z)} \quad , \qquad P_N(Z) = \sum_{i=0}^{N} c_i^{(N)}\,Z^i \ ,
\end{equation}
where $P_N$ is a polynomial of degree $N$ in $Z$ with constant coefficients $c_i^{(N)}=c_i^{(N)}(\alpha_1,\alpha_2,\beta,m_1,m_2,M,e_1,e_2)$. The explicit form of these coefficients, in the case $m_1=m_2=m,\,e_1=e_2=e$, is presented in the Appendix.
\vspace{0.2cm}

In particular, for the Helium atom ($m_1=m_2=1\,$, $e_1=e_2=-1$,$\,Z=2$) with finite nuclear mass $M=7294.261824 \,a.u.$, minimization of $E_{\rm m,var}^{(3D)}$ with $\alpha_1=\alpha_2$ gives the value
\[
E_{\rm m,var I}^{(3D)}(Z=2) \ = \ -2.8891879 \,a.u. \ ,
\]
at $\alpha_1=\alpha_2=0.92887416,\,\beta=-0.2546058$. Releasing the condition $\alpha_1=\alpha_2$, the variational energy turns out to be
\[
E_{\rm m,var II}^{(3D)}(Z=2) \ = \ -2.8991105 \,a.u. \ ,
\]
at $\alpha_1=1.1031235,\,\alpha_2=0.72010632,\,\beta=-0.207181825$. They have to be compared with the \emph{exact} result \cite{Duan}:
\[
E_{\rm m,exact}^{(3D)}(Z=2) \ = \ -2.9033045\,a.u.    \ .
\]
Since $E_{\rm m,var}^{(3D)}$ admits an analytical expression, it is straightforward to study different systems. In particular, in Table \ref{ExAt}
we show the comparison of the ground-state energy obtained in the present study with the results calculated in \cite{Korobov} where a large  number ($\sim 1000$) of basis functions were employed.

\vspace{0.2cm}

\begin{table}[h]
\[\begin{array}{|c||c|c|c|c|c|}
\hline
  {\rm System} & \hspace{0.5cm} E_0\,(a.u.) \hspace{0.5cm} & \hspace{0.5cm} E_0^{(a)}\,(a.u.) \hspace{0.5cm} & \hspace{0.5cm} \alpha_1 \hspace{0.5cm} & \hspace{0.5cm} \alpha_2 \hspace{0.5cm} & \hspace{0.5cm} \beta  \\ \hline
  e^-\,e^-\,e^+ & -0.256692 & -0.262005 & 0.520138 & 0.147915 & -0.005991 \\
  \hline
  H_2^+         & -0.592720 & -0.597139 & 0.333979 & 0.786127 & 0.445961 \\
  \hline
  H^-           & -0.523557 & -0.527751 & 1.0743677 & 0.4831917 & -0.146303 \\
  \hline
\end{array}\]
\caption{Comparison of the ground-state energy of the positive
hydrogen molecular ion and the system $e^-\,e^-\,e^+$ obtained in this work ($E_0$) with the theoretical
calculations ($ E_0^{(a)} $) presented in \cite{Korobov}. Similarly, for the ion $H^-$ we compare our result with those presented in \cite{Korobov2}. }
\label{ExAt}
\end{table}

In Table \ref{T3}, we display the ground state energy $E_{\rm m,var}^{(3D)}\equiv E_0$ of the 3D Helium atom and its isoelectronic ions for $Z = 2,\ldots, 10$. Our variational results are compared with the highly accurate values ($\sim 35$ significant digits) obtained in \cite{Nakashima}. The corresponding relative error is presented as well. At $Z=2$, this error turns out to be $\sim 0.001$ and it decreases monotonically up to $0.00004$ at $Z=10$. It exhibits the high quality of the Hylleraas trial function (\ref{psi1m}).

\begin{center}
\begin{figure}[h]
\includegraphics[scale=0.5]{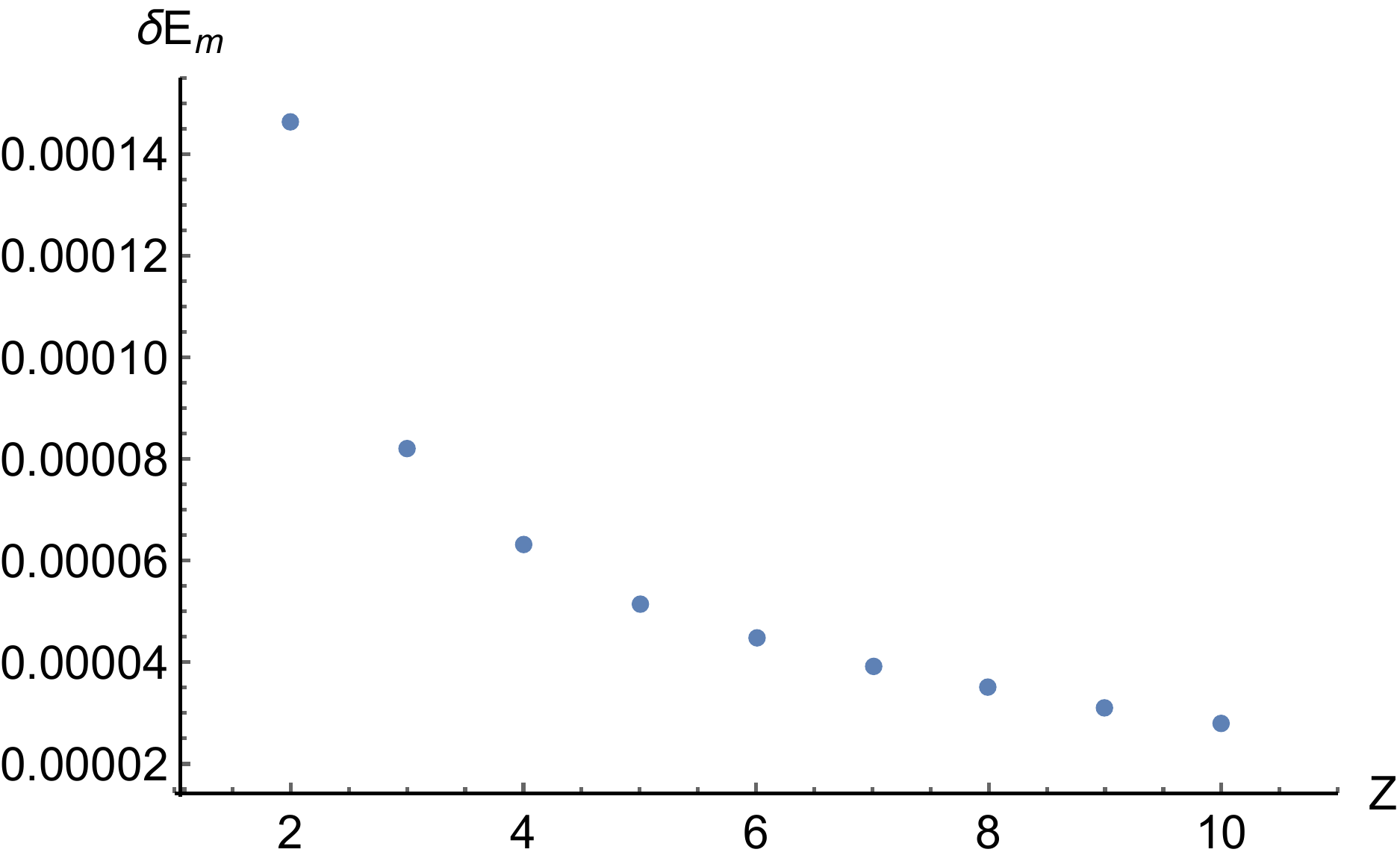}
\caption{The ground state energy of 3D Helium atom and its isoelectronic ions for $Z = 2,3,\ldots, 10$. The relative difference $\delta E_m=\frac{|E_{0,\rm finite}-E_{0,\rm infinite}|}{|E_{0,\rm finite}|}$ for $2\leq Z \leq 10$. The energies were calculated from the 3-parametric Hylleraas-type function (\ref{psi1m}).}
\label{T4}
\end{figure}
\end{center}

Finally, in Fig. \ref{T4} we present the corresponding comparison between the finite and infinite nuclear mass cases within the variational approach.

\begin{table}[h]
\[\begin{array}{|c|c|c|c|c|c|c|}
\hline
\hspace{0.3cm} Z \hspace{0.3cm} & \hspace{0.3cm} E_0\,(a.u.) \hspace{0.3cm} & \hspace{0.3cm} E_{0,{\rm exact}}\,(a.u.) \hspace{0.3cm} & \hspace{0.4cm} \Delta\,E \hspace{0.3cm} & \hspace{0.4cm}  \alpha_1 \hspace{0.4cm} &\hspace{0.4cm}  \alpha_2 \hspace{0.4cm} & \hspace{0.4cm} \beta \hspace{0.4cm} \\
\hline
\hline
2 & -2.89911 & -2.903304  &   0.001444 & 1.103123 &  0.720106 &  -0.207182 \\
3 & -7.27511 & -7.279321 &    0.000578 & 1.099262 &  0.787267 &  -0.220592 \\
4 & -13.65050 & -13.654709 &  0.000308 & 1.093341 &  0.822938 &  -0.226777 \\
5 & -22.02564 & -22.029846 &  0.000191 & 1.087862 &  0.845780 &  -0.230357 \\
6 & -32.40054 & -32.404733  & 0.000129 & 1.083099 &  0.861923 &  -0.232648 \\
7 & -44.77546 & -44.779658  & 0.000094 & 1.078982 &  0.874092 &  -0.234331 \\
8 & -59.15034 & -59.154533  & 0.000071 & 1.075412 &  0.883655 &  -0.235556 \\
9 & -75.52531 &  -75.529499 & 0.000055 & 1.072286 &  0.891420 &  -0.236509 \\
10& -93.90001 & -93.904195 &  0.000044 & 1.069102 &  0.898483 &  -0.238653 \\
\hline
\end{array}\]
\caption{The ground state energy $E_0$ of 3D Helium atom and its isoelectronic ions for $Z = 2,3,\ldots, 10$ with finite nuclear mass. The values of masses were taken from (\cite{Nakashima}). The second column represents the variational energies from the 3-parametric trial function (\ref{psi1m}). The third column corresponds to the exact numerical values (see \cite{TVH,Nakashima}). The relative error $\Delta\,E=\frac{|E_{0,{\rm exact}}-E_0|}{|E_{0,{\rm exact}}|}$ is displayed as well. The last three columns show the optimal variational parameters in (\ref{psi1m}). They are smooth, slow-changing functions of the nuclear charge $Z$.}
\label{T3}
\end{table}

\clearpage

\section{Conclusions}

Summarizing, for Helium-like ions in $d-$dimensions an approximate compact analytical expression for the ground state energy of the system is constructed in the non-relativistic domain $Z\leq 10$ for the lowest values $d=2,3,4,5$. We call them generalized Majorana solutions. The basic building block is a Hylleraas-type trial function given by a 2-parametric correlated exponential function in the Hylleraas variables ($r_1,r_2,r_{12}$). In particular, by pure algebraic means these generalized Majorana solutions allow us to calculate the (first) critical charge with reasonable accuracy and they provide a good estimation of the leading terms in the celebrated $\frac{1}{Z}$ expansion. At $d=3$, a similar situation occurs for the Puiseux expansion.

Especially, it was demonstrated that for odd $d\geq 3$ the Majorana solution is given by a rational algebraic function of the $Z$ whilst for even $d\geq 2$ it turns out to be a more complicated non-algebraic expression. In the latter case it is quite a non-trivial task due to the elaborate form of the Jacobian in the functional energy. To the authors knowledge, the corresponding 2D Majorana solution is reported for the first time.
The key element was the introduction, in the 3-dimensional $r-$space, of a nonlinear change of variables.

Originally, in the computations reported here the nuclear charge was assumed to be of infinite mass. This assumption, though unnecessary, was
adopted to facilitate the computations and the comparison with earlier results by other authors. At $d=3$, an improved 3-parametric correlated Hylleraas-type trial was used to compute the finite mass effects in the generalized Majorana solution for a three-body Coulomb system with arbitrary charges and masses. Similar to the infinite nuclear mass case\cite{HOAT}, it leads to surprisingly accurate values where the relative error smoothly decreases from $10^{-3}$ at $Z=2$ up to $10^{-5}$ at $Z=10$. In the present study concrete results, beyond the Born-Oppenheimer approximation, for the systems $e^-\,e^-\,e^+$, $H_2^+$ and $H^-$ were calculated explicitly.

We also showed that the Shannon entropy for Helium-like systems exhibits the following behavior: i) for a fixed value of $d$, the Shannon entropy is a monotonic decreasing function of the nuclear charge $Z$, this is explained by the fact that the larger the $Z$, the greater the Coulomb attraction between the electrons and the nucleus is, hence, the electrons are located at a smaller distance from the nucleus (localization), and ii) for a fixed value of the nuclear charge $Z$ the Shannon entropy grows with dimensionality, a similar situation occurs for the hydrogen atom in $d$-dimensions.

Finally, when the hydrogen-like atoms are confined inside an impenetrable spherical box of radius $R_c$, the relativistic effects become more significant, especially in the strong confinement regime. It is important to know the value of relativistic corrections such as the kinetic energy correction, spin-orbit coupling and the one-body and two-body Darwin terms, as a function of the confinement radius $R_c$, nuclear charge $Z$ and the space dimensionality $d$.

Quantum entanglement in the ground state of the helium atom has been studied in three dimensions by means of von Neumann entropy and linear entropy. It is interesting to explore entanglement in the helium atom in $d$-dimensions.

\vspace{2.0cm}

{\bf Acknowledgments}

\vspace{2.0cm}

One of the authors (HOP) thanks the support through the Programa Especial de Apoyo a la Investigaci\'on 2019, Universidad Aut\'onoma Metropolitana, (I30). AMER thanks the support through the Programa Especial de Apoyo a la Investigaci\'on 2021, UAM-I.

\vspace{3.0cm}

\section{DATA AVAILABILITY}

The authors confirm that the data supporting the findings of this study are available within the article.

\clearpage

\appendix
\section{Coefficients $c_i^{(N)}$}

For $m_1=m_2=m,\,e_1=e_2=e$, the coefficients $c_i^{(N)}$ appearing in $E_{\rm m,var}^{(3D)}$ (\ref{Em3Dv}). For the numerator $P_{10}(Z)=\sum_{i=0}^{10} c_i^{(10)}\,Z^i$ we obtain
{\small
\begin{equation}\label{}
\begin{aligned}
& c_{10}^{(10)} \ = \ \alpha _s^2 \,\bigg[\alpha _s^6 \left(\alpha _s \left(2\, e\, m\, M+(m+M)\, \alpha _s\right)-2 \,(m+M)\, \alpha _p\right)+128 \alpha _p^3  \,e\, m\, M\, \alpha _s+128\, \alpha _p^4 \, (m+M)\bigg]
\\ &
\vspace{0.15cm}
c_{9}^{(10)} \ = \  \alpha _s \,\bigg[\,\alpha _s^3\, \big(\,\alpha _p \,\alpha _s^2\, [\alpha _s \left(2 e^2 m M-23 \beta  m-21 \beta  M\right)+4 \beta  e m M\,]+2 M \alpha _p^2 \left(\alpha _s \left(\beta +e^2 m\right)+192 \beta  e m\right)
\\ &
+\beta  \alpha _s^4 \left(28 e m M+13 (m+M) \alpha _s\right)\,\big)+8 \alpha _p^3 \, \alpha _s \,[\alpha _s \left(5 e^2 m M+58 \beta  (m+M)\right)+40 \beta  e m M\,]+160 \alpha _p^4 \,\beta\,  M\,\bigg]
\\ &
\vspace{0.15cm}
c_{8}^{(10)} \ = \  \beta  \bigg[\alpha _s^3 \big(2 \alpha _p \alpha _s^2 [\,\alpha _s \left(13 e^2 m M-55 \beta  m-42 \beta  M\right)+212 \beta  e m M\,]+4 \alpha _p^2\, (\alpha _s \left(35 e^2 m M+156 \beta  m+161 \beta  M\right)
\\ &
\ + \ 336\, \beta\,  e\, m\, M\,)+\alpha _s^4 \left(\alpha _s \left(2\, e^2\, m\, M+73\, \beta \, m+75 \,\beta\,  M\right)+172 \,\beta\,  e\, m\, M\,\right)\big)
\\ &
\ + \ 16\, \alpha _p^3 \,\alpha _s \,[\alpha _s \left(9 e^2 m M+26 \beta  m+64 \beta  M\right)+8\, \beta \, e \,m\, M\,]\ + \ 64\, \alpha _p^4 \, \beta\,  M\,\bigg]
\\ &
\vspace{0.15cm}
c_{7}^{(10)} \ = \  \beta ^2 \,\alpha _s\, \bigg[\,32\, \alpha _p^3 \,M \,\left(26 \beta +5 e^2 m\right)+\alpha _s \bigg(8 \alpha _p \alpha _s^2 \left(\alpha _s \left(33\, e^2\, m \,M+11 \beta  m+29 \beta  M\right)+236 \beta \, e\, m\, M\,\right)
\\ &
\ + \ 8 \,\alpha _p^2\, \big[\alpha _s \,\left(79\, e^2\, m\, M+138 \,\beta\,  m+256\, \beta\,  M\right)+168 \,\beta\,  e \,m\, M\,\big]
\\ &
\ + \ \alpha _s^4 \left(\alpha _s \left(26 \,e^2\, m\, M+231 \,\beta\,  m+257\, \beta\,  M\right)+732 \,\beta\,  e\, m\, M\,\right)\bigg)\bigg]
\\ &
\vspace{0.15cm}
c_{6}^{(10)} \ = \  2 \,\beta ^3\, \bigg[32\, \alpha _p^3 \, M \,(4\, \beta +e^2\, m\,)+\alpha _s \bigg(4 \alpha _p \alpha _s^2 \left(\alpha _s [139 \,e^2\, m\, M+70 \beta\,  m+193 \,\beta \, M]+376\, \beta \, e\, m\, M\right)
\\ &
\ + \ 8 \,\alpha _p^2\, \left(\alpha _s \,\left(67 \,e^2\, m\, M+30\, \beta\,  m+166\, \beta\,  M\,\right)+24 \,\beta\,  e\, m\, M\right)
\\ &
\ + \ \alpha _s^4\, \left(\alpha _s \,\left(93 \,e^2\, m\, M+265\, \beta\,  m+338 \,\beta\,  M\right)+1012\, \beta\,  e\, m\, M\,\right)\bigg)\bigg]
\\ &
\vspace{0.15cm}
c_{5}^{(10)} \ = \  2 \,\beta ^4 \,\alpha _s \,\bigg[16 M \alpha _p^2 \left(53 \beta +26 e^2 m\right)+4 \alpha _p \alpha _s \big(\alpha _s [283\, e^2\, m\, M+64 \beta  m+388 \,\beta \, M\,]+256\, \beta\,  e\, m\, M\,\big)
\\ &
\ + \ \alpha _s^3 \,[\,\alpha _s\, \left(363 \,e^2\, m\, M+412\, \beta \, m+707\, \beta\,  M\,\right)+1648\, \beta\,  e\, m\, M\,]\,\bigg]
\\ &
\vspace{0.15cm}
c_{4}^{(10)} \ = \  4 \,\beta ^5 \,\bigg[16\, M\, \alpha _p^2 \left(7 \beta +4 e^2 m\right)+4 \alpha _p \alpha _s \,\big(\,\alpha _s \left(153 e^2 m\, M+8\, \beta \, (m+25\, M)\,\right)+32\, \beta\,  e\, m\, M\,\big)
\\ &
\ + \ \alpha _s^3\, \big(\,\alpha _s \,\left(403\, e^2\, m\, M+188\, \beta\,  m+549\, \beta\,  M\right)+752\, \beta\,  e\, m\, M\,\big)\,\bigg]
\\ &
\vspace{0.15cm}
c_{3}^{(10)} \ = \  8 \beta ^6 \alpha _s \bigg[\,4 \,M \,\alpha _p \,\left(54 \beta +43 e^2 m\right)+\alpha _s \,\big(\,\alpha _s \left(265\, e^2\, m\, M+44 \beta  m+298 \beta  M\right)+176 \beta\,  e\, m\, M\big)\,\bigg]
\\ &
\vspace{0.15cm}
c_{2}^{(10)} \ = \  16 \,\beta ^7\, \bigg[4 \,M \,\alpha _p\, \left(6 \,\beta +5 \,e^2\, m\right)+\alpha _s \big(\alpha _s \left(103 \,e^2\, m \,M+4 \beta  m+106 \,\beta\,  M\right)+16 \,\beta \, e \,m \,M\,\big)\,\bigg]
\\ &
\vspace{0.15cm}
c_{1}^{(10)} \ = \ 704 \,\beta ^8\, M\, \alpha _s \,\left(\,\beta \,+\,e^2 \,m\,\right)
\\ &
\vspace{0.15cm}
c_{0}^{(10)} \ = \ 128\, \beta ^9\, M \,\left(\,\beta \,+\,e^2\, m\,\right)
\end{aligned}
\end{equation}
where $\alpha _s \ \equiv \ \alpha_1 \ + \ \alpha_2 \quad ; \qquad \alpha _p \ \equiv \ \alpha_1 \, \alpha_2 \ $.

}

{\small

As for the denominator $P_{8}(Z)=\sum_{i=0}^{8} c_i^{(8)}\,Z^i$, the coefficients take the form

\begin{equation}\label{}
\begin{aligned}
& c_{8}^{(8)} \ = \ 2\, m \,M\, \alpha _s^2 \,\left(\,64\, \alpha _p^3 \ + \ \alpha _s^6\,\right)
\\
& c_{7}^{(8)} \ = \ 2 \,\beta \, m\, M\, \alpha _s\, \left(80\, \alpha _p^3\ + \ \alpha _p \,\alpha _s^4 \ + \ 192 \,\alpha _p^2 \,\alpha _s^2 \ + \ 13\, \alpha _s^6\,\right)
\\
& c_{6}^{(8)} \ = \  2 \,\beta ^2\, m\, M\, \left(\,32\, \alpha _p^3\ + \ 202\, \alpha _p\, \alpha _s^4\ + \ 432\, \alpha _p^2 \,\alpha _s^2 \ + \ 73\, \alpha _s^6\,\right)
\\
& c_{5}^{(8)} \ = \  2\, \beta ^3\, m\, M\, \alpha _s\, \left(336\, \alpha _p^2\ + \ 664\, \alpha _p\, \alpha _s^2\ + \ 295\, \alpha _s^4\,\right)
\\
& c_{4}^{(8)} \ = \  4 \,\beta ^4 \,m \,M \,\left(\,48\, \alpha _p^2\ + \ 424\, \alpha _p\, \alpha _s^2 \ +\  361\, \alpha _s^4\,\right)
\\
& c_{3}^{(8)} \ = \  16 \,\beta ^5 \,m\, M\, \alpha _s\, \left(\,64\, \alpha _p \ + \ 127\, \alpha _s^2\,\right)
\\
& c_{2}^{(8)} \ = \  32 \,\beta ^6 \,m\, M\, \left(8\, \alpha _p\ + \ 51\, \alpha _s^2\,\right)
\\
& c_{1}^{(8)} \ = \  704\, \beta ^7\, m\, M\, \alpha _s
\\
& c_{0}^{(8)} \ = \  128 \,\beta ^8\, m\, M \ .
\end{aligned}
\end{equation}

}

\end{document}